\documentclass[11pt]{article}\usepackage[]{graphicx}\usepackage[]{xcolor}
\makeatletter
\def\maxwidth{ %
  \ifdim\Gin@nat@width>\linewidth
    \linewidth
  \else
    \Gin@nat@width
  \fi
}
\makeatother

\definecolor{fgcolor}{rgb}{0.345, 0.345, 0.345}

\usepackage{framed}
\makeatletter
 {\par\unskip\endMakeFramed%
 \at@end@of@kframe}
\makeatother

\definecolor{shadecolor}{rgb}{.97, .97, .97}
\definecolor{messagecolor}{rgb}{0, 0, 0}
\definecolor{warningcolor}{rgb}{1, 0, 1}
\definecolor{errorcolor}{rgb}{1, 0, 0}
\newenvironment{knitrout}{}{} 

\usepackage{alltt}
\usepackage[margin=1in]{geometry}
\usepackage{amsmath, amsfonts, amsthm}
\usepackage{natbib}
\usepackage{hyperref}
\usepackage{afterpage}
\usepackage{pdflscape}
\usepackage{csquotes}
\usepackage{tabularx}
\usepackage{booktabs}
\usepackage[figuresright]{rotating}
\usepackage{newtxtext}
\usepackage[subscriptcorrection]{newtxmath}
\usepackage{authblk}

\newcommand{\hti}{{\hat{\theta}_i}}

\newcommand{\ti}{{\theta_i}}
\newcommand{\hmu}{{\hat{\mu}}}
\newcommand{\hmuivw}{{\hat{\mu}_{\scriptscriptstyle{\text{IVW}}}}}
\newcommand{\hsi}{{\hat{\sigma}^2_i}}
\newcommand{\si}{{\sigma^2_i}}
\newcommand{\taus}{{\tau^2}}
\newcommand{\htaus}{{\hat{\tau}^2}}
\newcommand{\sima}{\mathrel{\overset{\text{a}}{\thicksim}}} 
\newcommand{\wit}{{w_i(\taus)}}
\renewcommand{\d}{{\mathrm{d}}}
\newcommand{\remltaus}{{\hat{\tau}^2_{\scriptscriptstyle{\text{REML}}}}}

\DeclareMathOperator{\se}{se}  
 
\DeclareMathOperator{\Q}{Q}

\DeclareMathOperator{\Nor}{N} 
\DeclareMathOperator{\Chi}{\chi^2}

\DeclareMathOperator{\SN}{SN}
\DeclareMathOperator{\Uni}{U} 

\newcommand{\taustar}{{{\tau}_{(b)}^{2*}}}
\newcommand{\mustar}{{{\mu}_{(b)}^{*}}}

\newcommand{\A}{{\mathbf{A}}}
\newcommand{\B}{{\mathbf{B}}}

\newcommand{\niter}{{n_{\text{sim}}}}

\newcommand{\floor}[1]{\lfloor #1 \rfloor}

\newtheorem{condition}{Condition}[section]
\newtheorem{definition}{Definition}[section]
\newtheorem{assumption}{Assumption}[section]
\newtheorem{step}{Step}[section]

\IfFileExists{upquote.sty}{\usepackage{upquote}}{}
\begin{document}
\title{\textbf{Edgington's Method for Random-Effects Meta-Analysis Part I: Estimation}}

\author[1]{David Kronthaler\thanks{david.kronthaler@uzh.ch}}
\author[1]{Leonhard Held\thanks{leonhard.held@uzh.ch}}

\affil[1]{Epidemiology, Biostatistics and Prevention Institute, Department of Biostatistics, University of Zurich, Switzerland}

\date{} 
\maketitle

\begin{framed}
\noindent\textbf{Note.} This preprint has been merged into arXiv:2510.13216, \emph{Prediction intervals for random-effects meta-analysis based on confidence distributions and Edgington's method}. Readers should consult that manuscript for the complete and current version of this work. The present preprint is retained for citation continuity.
\end{framed}

\section*{Abstract}
Meta-analysis can be formulated as combining $p$-values across studies into a joint $p$-value function, from which point estimates and confidence intervals can be derived. We extend the meta-analytic estimation framework based on combined $p$-value functions to incorporate uncertainty in heterogeneity estimation by employing a confidence distribution approach. Specifically, the confidence distribution of Edgington's method is adjusted according to the confidence distribution of the heterogeneity parameter constructed from the generalized heterogeneity statistic. Simulation results suggest that 95\% confidence intervals approach nominal coverage under most scenarios involving more than three studies and heterogeneity. Under no heterogeneity or for only three studies, the confidence interval typically overcovers, but is often narrower than the Hartung--Knapp--Sidik--Jonkman interval. The point estimator exhibits small bias under model misspecification and moderate to large heterogeneity. Edgington's method provides a practical alternative to classical approaches, with adjustment for heterogeneity estimation uncertainty often improving confidence interval coverage.\\[0.5ex]
\noindent\textbf{Keywords}: 
confidence distribution; estimation uncertainty; heterogeneity (variance); meta-analysis, $p$-value function.

\maketitle

\newpage

\section{Introduction}

Meta-analysis is a statistical method for quantitatively synthesizing evidence from independent studies \citep{borenstein2021introduction}. Compared to individual studies, systematic reviews and meta-analyses provide stronger evidence and often inform clinical guidelines and policy decisions \citep{walker2008meta}. Classical meta-analysis typically employs a weighted average of effect estimates, most commonly using inverse-variance weighting (IVW). Under a random-effects model, accounting for systematic differences between studies through between-study heterogeneity, the variances of effect estimates are additively increased by the estimated variance of true effects \citep{DerSimonian1986}. Under exchangeability or random sampling of true effects, the resulting estimator is consistent and efficient for the mean of the underlying effect distribution \citep{borenstein2021introduction}, with several confidence intervals proposed \citep{Hartung2001, sidik2002simple, henmi2010confidence}. A limitation of these intervals is their symmetric form, which fails to capture data skewness.

Meta-analytic estimation can be reformulated within the general framework of combining $p$-values across studies using $p$-value functions \citep{Fraser2019, infanger2019p}, or equivalently, confidence distributions \citep{SCHWEDER2002, xie2011confidence, Marschner2024}. Specifically, a weighted Stouffer method recovers the classical IVW estimator \citep{Senn2021}. Recently, \citet{heldpawelhofman2024} suggested that alternative $p$-value combination methods may serve as substitutes or complements to the classical approach. Several of these yield confidence intervals that are not constrained to symmetry, a desirable property, as skewed distributions of effect estimates should be reflected in the statistical inference, rather than being simplified into symmetric summaries. This aligns with increasing calls for meta-analytic methods that appropriately handle skewed data \citep{higgins2008meta, yang2016meta, noma2022meta}.

Among the investigated $p$-value combination methods, Edgington’s approach \citep{edgington1972additive} based on the sum of $p$-values has been recommended due to its invariance to the orientation of the alternative under which one-sided $p$-values are constructed and its ability to produce virtually unbiased point estimates and confidence intervals with near-nominal coverage. A limitation of the proposed procedure is that it does not account for uncertainty in the estimation of between-study heterogeneity. This is particularly relevant for meta-analyses based on a small number of studies, where heterogeneity is estimated with low precision. For example, the method by \citet{Hartung2001} and \citet{sidik2002simple} accounts for this by applying a $t$-distribution, typically yielding better coverage than the classical random-effects interval \citep{borenstein2021introduction}. Generally, many inferential procedures in meta-analysis benefit from incorporating parameter estimation uncertainty. For instance, prediction intervals adjusted in this way \citep{Higgins2008, Partlett2016} usually outperform those that ignore such uncertainty \citep{Skipka2006}.

To this end, we propose an extension of Edgington’s method that incorporates heterogeneity estimation uncertainty. Our approach adjusts the combined $p$-value function, respectively confidence distribution, according to a confidence distribution of the heterogeneity parameter, implied by the generalized heterogeneity statistic \citep{Viechtbauer2006}. While we illustrate this using Edgington’s method, the approach is applicable to other $p$-value combination methods (e.g., \citeauthor{tippett1931methods}, \citeyear{tippett1931methods}; 
\citeauthor{pearson1933method}, \citeyear{pearson1933method}; 
\citeauthor{fisher1934statistical}, \citeyear{fisher1934statistical}; 
\citeauthor{wilkinson1951statistical}, \citeyear{wilkinson1951statistical}) that yield a valid confidence distribution of the average effect. However, in the simulation study by \citet{heldpawelhofman2024}, these methods showed relatively poor performance, and it remains up to investigation whether such an adjustment could improve inference.

In part one of this series, we focus on the target of estimation; in part two, we extend the methodology to the prediction of future study effects. The remainder of this article is structured as follows: we first introduce the general approach to meta-analysis via $p$-value combination, then present methods to incorporate uncertainty about the heterogeneity estimate in the estimation of the average effect. We illustrate the methods using a case study, followed by an evaluation of its performance in a simulation study.

\section{Methods}

\subsection{Random-Effects Meta-Analysis}

The presented methods are developed within the random-effects framework, which assumes that the true effects $\ti$, $i \in \{1,\ldots, k\}$, from $k$ observed studies are exchangeable and jointly normal distributed. Variability in effect estimates $\hti$ is attributed to sampling noise and between-study heterogeneity. The random-effects model can be formulated as a  hierarchical $k + 2$ parameter model, which collapses to a two-parameter model by marginalization:
\begin{equation}\label{eq:rema}
\ti  \sim \Nor(\mu, \taus), \quad \hti\mid\ti \sim \Nor(\ti, \hsi), \quad \Rightarrow \quad \hti \sim \Nor(\mu, \taus + \hsi).
\end{equation}
In this model, $\mu$ represents the average effect across studies, $\taus$ quantifies the variance of true effects, and $\hsi$ denotes the squared standard error from study $i$, which is treated as fixed. Approximate normality of $\hti$ is commonly justified by the Central Limit Theorem (CLT), provided sufficiently large study sample sizes \citep{rice2018re}. Often, transformed estimates $h(\hti)$, where $h(\cdot)$ maps effects to a scale closer to normality, can be useful: for example, applying the logit to probabilities or converting correlations to Fisher $Z$-scores \citep{schwarzer2021meta, field2010meta}. Normality of $\ti$ is often simplistically assumed \citep{higgins2008meta}, but it is commonly acknowledged that location estimates remain fairly robust to misspecification of the random-effects distribution \citep{Lee2007}. 

The presented methods are not applicable under a fixed-effect (or common-effect) framework, since they explicitly account for heterogeneity through a confidence distribution that always assigns non-zero mass to heterogeneity greater than zero. The fixed-effects framework \citep{rice2018re} is also inapplicable, as the approach assumes exchangeability of true effects.

\subsection{$P$-Value Functions and Confidence Distributions}

\begin{figure}
\centering
\begin{knitrout}
\definecolor{shadecolor}{rgb}{0.969, 0.969, 0.969}\color{fgcolor}

{\centering \includegraphics[width=1\linewidth]{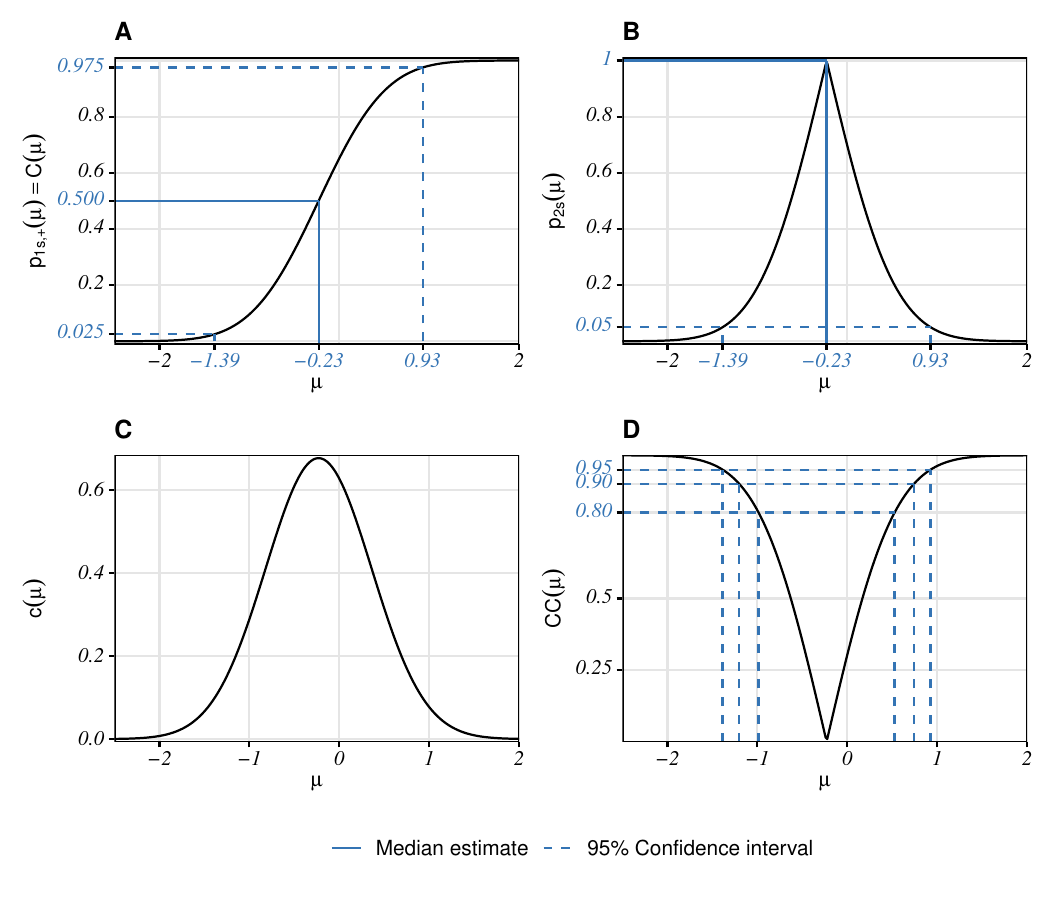} 

}

\end{knitrout}
\caption{Results from \citet{glemain2002tamsulosin}, see Table~\ref{tab:serenoadata}. Wald test for $\mu$ with $\hmu = -0.23$ and $\se(\hmu) = 0.59$: (A) One-sided $p$-value function for the alternative "greater", corresponding to the confidence distribution function; (B) two-sided $p$-value function; (C) confidence density; (D) confidence curve.}
\label{fig:pfunexample}
\end{figure}

A $p$-value function treats the $p$-value as a function of the parameter of interest, providing evidence against all possible null hypotheses and supporting point and interval estimation. Consider the Wald test, which is later also used for meta-analysis estimation, for a parameter $\mu$ with estimator $\hmu$ and standard normal pivot $Z(\mu) = (\hmu - \mu)/\se(\hmu)$. The corresponding one-sided (1s) and two-sided (2s) $p$-value functions are
\begin{align*}
p_{\text{1s},+}(\mu) &= 1 - \Phi\left(Z(\mu)\right) && \text{for the alternative "greater"}, \\
p_{\text{1s},-}(\mu) &= \Phi\left(Z(\mu)\right) && \text{for the alternative "less"}, \\
p_{\text{2s}}(\mu) &= 2 \min\left\{p_{\text{1s},+}(\mu),\; p_{\text{1s},-}(\mu)\right\},
\end{align*}
where $\Phi(\cdot)$ denotes the standard normal cumulative distribution function (CDF). Figures~\ref{fig:pfunexample}A and~\ref{fig:pfunexample}B show the one-sided $p$-value function for the "greater" alternative and the two-sided $p$-value function for the Wald test with $\hmu = -0.23$ and $\se(\hmu) = 0.59$. These estimates, reported by \citet{glemain2002tamsulosin}, concern the effect of treatment with \textit{Serenoa repens} on prostate symptom scores and are included in the meta-analysis case study described in Section~\ref{sec:example}.

The $p$-value function has several direct applications: The value $\hmu = p_{\text{1s},+}^{-1}(0.5) = p_{\text{1s},-}^{-1}(0.5) = -0.23$ is the median estimate for $\mu$. The median estimate can also be obtained from the two-sided $p$-value function as the value of $\mu$ maximizing said function. A two-sided 95\% confidence interval for $\mu$ reaches from $p_{\text{1s},+}^{-1}(0.025) = p_{\text{1s},-}^{-1}(0.975) = -1.39$ to $p_{\text{1s},+}^{-1}(0.975) = p_{\text{1s},-}^{-1}(0.025) = $ 0.93. Further, one-sided $p$-value functions for the "greater" alternative are often the CDF of a confidence distribution.

Confidence distributions are frequentist probability distributions over the parameter space, constructed without invoking prior distributions \citep{Cox1958, nadarajah2015confidence, Marschner2024}. Rather than representing the inherent distribution of a parameter, they are modernly interpreted as sample-dependent distributional summaries of uncertainty \citep{Xie2013}. One interpretation sees them as encompassing all possible confidence intervals simultaneously \citep{SCHWEDER2002}, where the \emph{confidence} or \emph{confidence probability} assigned to a parameter subspace corresponds to the confidence level of the interval spanning it \citep{Marschner2024}. A formal definition of a sample-dependent confidence distribution is \citep{SCHWEDER2002}:

\begin{definition}[Confidence Distribution]\label{def:cd}
Let $\mathbf{Y}$ be a random vector with sample space $\mathcal{Y}$ and realization $\mathbf{y}$, and let $\mu \in \Theta$ be the parameter of interest. A function $C(\mathbf{Y}, \cdot): \Theta \rightarrow [0, 1]$ is called a \emph{confidence distribution} for $\mu$ if:

\begin{condition}
For each fixed $\mathbf{y} \in \mathcal{Y}$, $C(\mathbf{y}, \cdot)$ is a cumulative distribution function on $\Theta$.
\end{condition}

\begin{condition}
At the true parameter value $\mu = \mu_0$, $C(\mathbf{Y}, \mu_0)$ follows a standard uniform distribution: $C(\mathbf{Y}, \mu_0) \sim \Uni[0, 1]$.
\end{condition}

The function $C(\mathbf{Y}, \cdot)$ is called an \emph{asymptotic confidence distribution} if $C(\mathbf{Y}, \mu_0)$ converges in distribution to the standard uniform as the size of $\mathbf{Y}$ increases.
\end{definition}

The \emph{confidence density} is obtained by taking the derivative of the confidence distribution with respect to $\mu$:
    $$
    c(\mathbf{Y}, \mu) = \frac{\d C(\mathbf{Y}, \mu)}{\d\mu}.
    $$
The \emph{confidence curve} straightforwardly provides all two-sided confidence intervals across confidence levels \citep{birnbaum1961confidence}:
$$
CC(\mathbf{Y}, \mu) = \left| 1 - 2 C(\mathbf{Y}, \mu) \right|.
$$
For the example involving the Wald test, Figures~\ref{fig:pfunexample}A, \ref{fig:pfunexample}C and~\ref{fig:pfunexample}D also display the CDF, confidence density and confidence curve of the corresponding confidence distribution.

\subsection{Estimation in Meta-Analysis}\label{sec:estimation}

The marginal distributions $\hti \sim \Nor(\mu, \taus + \hsi)$ induce the $k$ pivots and corresponding one-sided $p$-value functions for the alternative "greater":
$$
Z_i(\mu) = \frac{\hti - \mu}{\sqrt{\taus + \hsi}} \sim \Nor(0, 1), \quad p_{\text{1s},+}(\mu) = 1 - \Phi\left(Z_i(\mu)\right),
$$
which are combined to yield a combined $p$-value function for $\mu$. One-sided $p$-value functions are preferred over two-sided $p$-value functions, since the latter may yield undesirable properties, such as poorly defined confidence intervals \citep{heldpawelhofman2024}.

Edgington's method of combining $p$-values corresponds to evaluating the CDF of the Irwin--Hall distribution ($F_{\text{IH}}$) with $k$ degrees of freedom at the sum of $p$-values:
$$
p_E(\mu) = F_{\text{IH}}(s) = \frac{1}{k!} \sum_{j = 0}^{\floor{s}} (-1)^j (s - j)^k, \quad s = \sum_{i = 1}^k p_i(\mu),
$$
where $\floor{s}$ denotes the lowest integer closest to $s$. For $k \ge 12$, Edgington's method is approximated using a normal distribution based on a CLT argument to mitigate overflow problems:
\begin{equation*}
p_E(\mu) = \begin{cases}
F_{\text{IH}}(s) & \text{if } k < 12 \\
\Phi\left(\sqrt{12k} \ (s/k - 1/2)\right) & \text{if } k \ge 12.
\end{cases}
\end{equation*}
Edgington's combined $p$-value function allows for the construction of point estimators and confidence intervals, similar to the Wald test above. 

The approach by \citet{heldpawelhofman2024} uses a plug-in estimate $\htaus$. By interpreting the combined $p$-value function as a confidence distribution, the method can be extended to incorporate uncertainty about heterogeneity estimation. Specifically, Edgington's method yields a confidence distribution of $\mu$, conditional on $\taus$, with density $c(\mu \mid \taus)$ (see the Supplementary Material for details). We propose to marginalize this confidence distribution by integrating over a confidence distribution of $\taus$ to account for uncertainty in heterogeneity estimation:
\begin{equation}\label{eq:adjustedcd}
c(\mu) =\int  c(\mu \mid \taus) \ c(\taus) \ \d\taus.
\end{equation}
Marginalizing a joint confidence distribution over a nuisance parameter is generally not guaranteed to yield a valid confidence distribution of the parameter of interest. Direct integration of the confidence density is typically only an approximation, whose accuracy should be assessed via simulation \citep{Schweder2016}. \citet{pawitan2021} show that under the normal model with pivots for location and scale, in that sense related to our meta-analysis setting, marginal confidence distributions obtained by integration can coincide with extended likelihood results. In generalized fiducial inference, closely related to confidence distributions, marginal fiducial distributions can be used for parameter-specific inference and often provide asymptotically correct coverage \citep{Hannig2014,murph2024generalized}. However, checking the approximation by simulation is still recommended.

%

A confidence distribution of $\taus$ is implied by an extension of Cochran’s $\Q$ statistic \citep{cochran1954combination}. Cochran’s $\Q$, commonly employed to test for heterogeneity \citep{Hoaglin2015}, can be generalized to depend on $\taus$, yielding the generalized heterogeneity statistic \citep{Viechtbauer2006}, also referred to as the $\Q$-profile heterogeneity statistic by \citet{Jackson2016}:
\begin{equation}\label{eq:genQ}
\Q(\taus) = \sum_{i=1}^k \frac{1}{\hsi + \taus} \ \left(\hti - \hmuivw(\taus)\right)^2.
\end{equation}
Here, the random-effects IVW estimator itself is a function of $\taus$. It was shown that $\Q(\taus)$ is distributed according to a $\Chi$-distribution with $k-1$ degrees of freedom under the model in~\eqref{eq:rema} \citep{Viechtbauer2006}, and it is therefore a pivotal statistic in $\taus$ \citep{Jackson2016} and yields a confidence distribution of $\taus$. Although this distribution is exact under the assumptions of~\eqref{eq:rema}, it relies on known within-study variances; in practice, the confidence distribution is hence only approximate.

Previously,~\eqref{eq:genQ} was introduced to derive a moment estimator of $\taus$, obtained by equating the statistic with its expected value $k-1$, referred to as the Paule--Mandel estimator \citep{paule1982consensus}. Using~\eqref{eq:genQ} for constructing confidence intervals for $\taus$, sometimes referred to as the $\Q$-profile method, was suggested by \citet{Viechtbauer2006} and is discussed by \citet{Jackson2016}. Treating~\eqref{eq:genQ} as a confidence distribution of $\taus$ can be viewed as an extension, representing the set of all possible confidence intervals derived via the $\Q$-profile method. We remark that \citet{dersimonian2007random} introduced an alternative generalization of Cochran's $\Q$ which relies on fixed weights and can also be used for confidence interval construction \citep{Jackson2013, Jackson2016}. Further, \citet{Nagashima2018} previously explored using a confidence distribution to account for uncertainty in heterogeneity estimation. They applied the exact confidence distribution of the standard $\Q$ statistic, derived by \citet{biggerstaff2008exact}, to incorporate this uncertainty into the construction of prediction intervals. However, they did not extend this approach to the generalized version that varies with $\taus$.

For the computation of~\eqref{eq:adjustedcd} we propose a Monte Carlo sampling algorithm. For each draw $b \in \{1,\ldots,B\}$, let $\taustar$ and $\mustar$ denote the sampled values. Each draw $b$ proceeds as follows:

\begin{enumerate}
\item Generate $\taustar$ by inverse transformation sampling \citep{ripley2009stochastic} from the confidence distribution of $\taus$ by numerically inverting $\Q(\taustar) = W_b$, where $W_b$ denotes a random variable from a $\chi_{k-1}^2$-distribution.
\item Generate $\mustar$ by inverse transformation sampling, exploiting that the confidence distribution of $\mu$ is conditional on $\taus$, by inverting $C(\mustar\mid\taustar) = U_b$, where $U_b$ is a standard uniform random variable.

\end{enumerate}
Samples $\tau_{(1)}^{2*},\dots, \tau_{(B)}^{2*}$ are independent by independence of $W_1,\dots,W_B$, and samples $\mu_{(1)}^{*}, \dots, \mu_{(B)}^{*}$ are independent by independence of $U_1,\dots,U_B$ and of $W_b$ with $U_b$. The empirical distribution of $\mustar$ estimates the marginalized confidence distribution in~\eqref{eq:adjustedcd}. Point estimates are taken as the mean of $\mu_{(1)}^{*}, \dots, \mu_{(B)}^{*}$, and confidence interval limits are obtained from sample quantiles. Alternatively, we explored deterministic integration by applying the change of variables formula to obtain the confidence density of $\taus$. Since $\Q(\taus)$ is monotonically decreasing in $\taus$ and its derivative is well-defined, and $\taus = \Q^{-1}(\Q(\taus))$, the confidence density of $\taus$ is
$$
c(\taus) = f_{\chi_{k-1}^2}(\Q(\taus)) \left| { \frac{\d \Q(\taus)}{\d \taus}} \right|,
$$
where $f_{\chi_{k-1}^2}(\cdot)$ denotes the density of a $\Chi$-distribution with $k-1$ degrees of freedom. The analytic derivative of $\Q(\taus)$ is provided in the Supplementary Material.  Then, the integral in~\eqref{eq:adjustedcd} is solved numerically using a global adaptive quadrature (GAQ) algorithm \citep{piessens2012quadpack, Rfntl}. Equi-tailed confidence intervals are derived by CDF inversion, while point estimates are computed by approximating the expected value by a weighted sum over midpoints using finite differences of the CDF. 

Table~S1 and Table~S2 display the results from a pilot simulation with 1000 iterations, varying the number of studies and heterogeneity under normally distributed true effects according to the simulation design presented in Section~\ref{sec:simstudy}. We found that the GAQ approach tends to produce slightly too wide marginal distributions and confidence intervals in scenarios with three or five studies. For ten or more studies, differences in confidence interval limits could be largely attributed to Monte Carlo error. Point estimates from both approaches were nearly identical. Although GAQ offers greater computational efficiency and avoids Monte Carlo noise, the results suggest numerical instability under scenarios with few studies, which may impact performance. Hence, we recommend using the Monte Carlo algorithm for computing confidence intervals, particularly in meta-analyses with few studies.

Both approaches allow to reconstruct the combined $p$-value function, since the tail mass of a confidence distribution at $\mu'$ equals the $p$-value of the one-sided test for $H_0: \mu = \mu'$ \citep{Xie2013}. For the Monte Carlo algorithm, this can be easily achieved by considering the empirical CDF, whereas the GAQ approach requires additional integration steps. The presented methods are implemented in the \textsf{edgemeta} package (\url{https://github.com/davidkronthaler-dk/edgemeta}).

\subsection{Example: \textit{Serenoa repens}}\label{sec:example}

\begin{figure}
\centering
\begin{knitrout}
\definecolor{shadecolor}{rgb}{0.969, 0.969, 0.969}\color{fgcolor}

{\centering \includegraphics[width=0.9\linewidth]{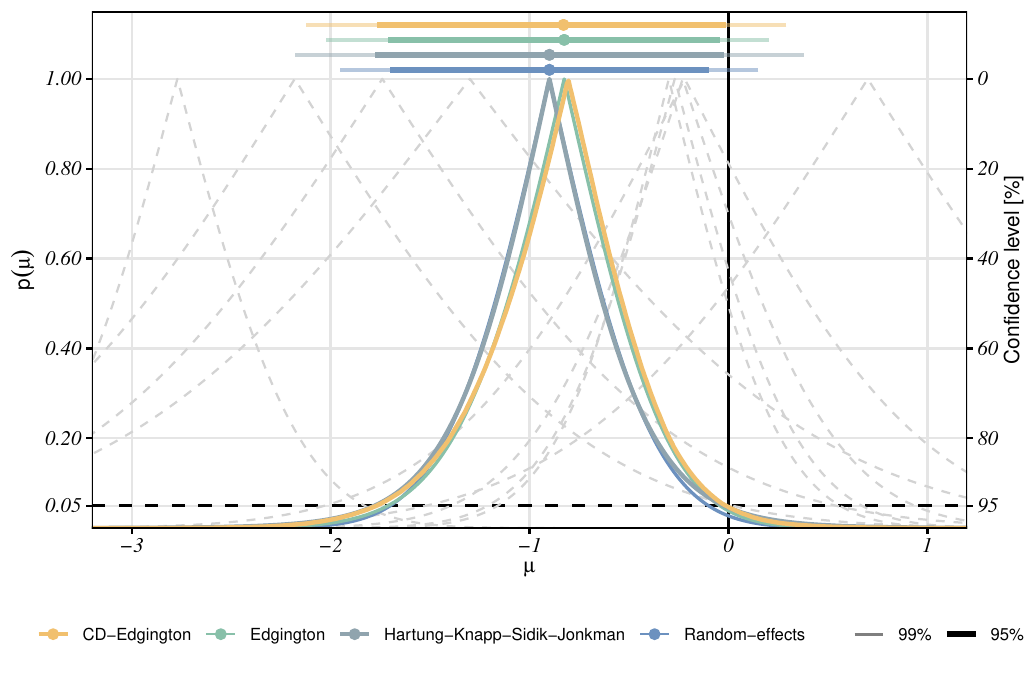} 

}

\end{knitrout}
\caption{Drapery plot displaying two-sided (combined) p-value functions from a random-effects meta-analysis of nine randomized controlled trials investigating \textit{Serenoa repens} for lower urinary tract symptoms. Confidence intervals at levels 95\% and 99\% are shown on top as telescope lines.}
\label{fig:serenoforest}
\end{figure}

\citet{Franco2023} present a meta-analysis of nine 1:1 randomized controlled trials investigating the effect of \textit{Serenoa repens} on lower urinary tract symptoms caused by benign prostatic enlargement, compared to placebo or no treatment. Effect measures are mean differences in International Prostate Symptom Scores at short-term follow-up (3 to 6 months), with lower values favoring treatment with \textit{Serenoa repens}. The data is summarized in Table~\ref{tab:serenoadata}. A drapery plot \citep{rucker2021beyond} displaying the results of a random-effects meta-analysis is shown in Figure~\ref{fig:serenoforest}. Estimators include the classical random-effects estimator, the Hartung--Knapp--Sidik--Jonkman (HKSJ) method, Edgington's method with additive heterogeneity adjustment using a fixed estimate $\htaus$ \citep{heldpawelhofman2024}, and the proposed CD-Edgington estimator.

\begin{figure}
\centering
\begin{knitrout}
\definecolor{shadecolor}{rgb}{0.969, 0.969, 0.969}\color{fgcolor}

{\centering \includegraphics[width=1\linewidth]{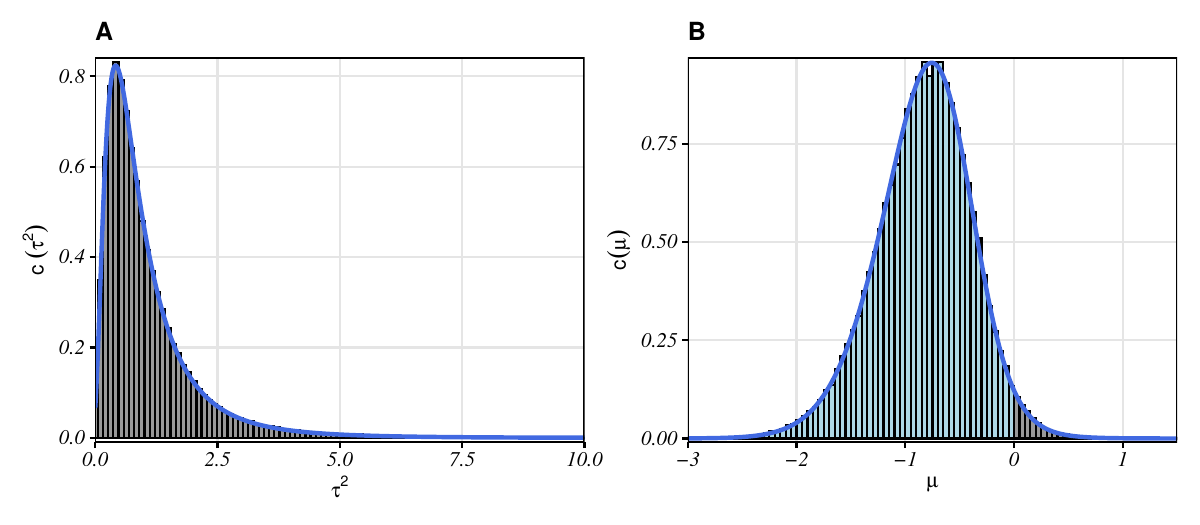} 

}

\end{knitrout}
\caption{Monte Carlo confidence distributions of: (A) the heterogeneity parameter $\taus$, shown with its analytical confidence density derived by change of variables; (B) the average effect $\mu$, shown with the confidence density computed via global adaptive quadrature integration. The blue colored area under the confidence density of the average effect corresponds to the confidence probability of this effect being smaller than zero.}
\label{fig:serenoaconfidences}
\end{figure}

Between-study heterogeneity is estimated as $\htaus$ = 0.85 (95\% confidence interval from 0.11 to 3.96; $p$ =  0.002; Higgins' $I^2$ of 67.4\%) based on restricted maximum likelihood (REML) estimation. The confidence distribution from the generalized heterogeneity statistic, both by Monte Carlo sampling and change of variables, is displayed in Figure~\ref{fig:serenoaconfidences}, together with the confidence distribution of the average effect, obtained from both Monte Carlo sampling and GAQ integration, providing a more complete presentation of parameter uncertainty beyond confidence intervals. For this example involving nine studies, both approaches produce virtually identical distributions. The confidence probability of the average effect being smaller than zero, naively interpreted as a beneficial effect on average, is 0.98, and corresponds to the area under the confidence density below zero, depicted in Figure~\ref{fig:serenoaconfidences}B in blue.

\begin{table}[ht]
\centering
\caption{Summary of \textit{Serenoa} studies analyzed in \citet{Franco2023}. All studies are 1:1 randomized controlled trials.} 
\label{tab:serenoadata}
\begin{tabular}{lrrrr}
 Study & N & Estimate & Standard error & 95\% CI \\ 
 Glemain (2002) & 329 & -0.23 & 0.59 & -1.40 to 0.94 \\ 
  Willetts (2003) & 93 & -1.74 & 1.16 & -4.02 to 0.54 \\ 
  Bent (2006) & 225 & -0.22 & 0.92 & -2.03 to 1.59 \\ 
  Shi (2008) & 94 & 0.70 & 1.13 & -1.52 to 2.92 \\ 
  Barry (2011) & 369 & -0.27 & 0.48 & -1.21 to 0.67 \\ 
  Gerber (2011) & 85 & -1.30 & 1.37 & -3.98 to 1.38 \\ 
  Argirovic (2013) & 199 & -0.30 & 0.44 & -1.16 to 0.56 \\ 
  Ye (2019) & 325 & -2.77 & 0.48 & -3.71 to -1.83 \\ 
  Sudeep (2020) & 99 & -2.18 & 1.12 & -4.38 to 0.02 \\ 
    
\multicolumn{5}{l}{\footnotesize CI = confidence interval.} \\ 
\end{tabular}
\end{table}

Table~\ref{tab:serepointestimates} provides a comparison of estimators for the average effect. The CD-Edgington estimator relates to the estimator by \citet{heldpawelhofman2024} as the HKSJ interval relates to the classical random-effects interval: both account for uncertainty in the estimation of heterogeneity, leading to wider confidence intervals. The skewness $\beta$ of confidence intervals is computed as \citep{Groeneveld1984}:
  $$
  \beta = \frac{\text{upper} + \text{lower} - \text{2 center}}{\text{upper} - \text{lower}}.
  $$
While classical random-effects and HKSJ confidence intervals are symmetric, both estimators based on Edgington's method reflect the left-skewed distribution of effect estimates (Fisher's weighted skewness coefficient \citep{ferschl1980deskriptive} of $-0.874$).

\begin{table}[ht]
\centering
\caption{Point estimates and 95\% confidence intervals for the average treatment effect across \textit{Serenoa} studies. $P$-values are computed for the null hypothesis $\mu_0 = 0$.} 
\label{tab:serepointestimates}
\begin{tabular}{lrrrrr}
 Estimator & Estimate & 95\% CI & Width & Skewness & p-value \\ 
 Random-effects & -0.90 & -1.70  to  -0.10 & 1.60 & 0.00 & 0.027 \\ 
  Hartung-Knapp-Sidik-Jonkman & -0.90 & -1.78  to  -0.02 & 1.76 & 0.00 & 0.046 \\ 
  Edgington & -0.83 & -1.71  to  -0.04 & 1.67 & -0.06 & 0.039 \\ 
  CD-Edgington (MC) & -0.83 & -1.77  to  -0.01 & 1.75 & -0.07 & 0.047 \\ 
  CD-Edgington (GAQ) & -0.83 & -1.76  to  -0.02 & 1.75 & -0.07 & 0.046 \\ 
    
\multicolumn{6}{l}{\footnotesize CI = confidence interval, GAQ = global adaptive quadrature, MC = Monte Carlo.} \\ 
\end{tabular}
\end{table}

\section{Simulation Study}\label{sec:simstudy}

\subsection{Design}
We present our proof-of-concept simulation study \citep{heinze2024phases} according to the ADEMP framework \citep{morris2019using}.

\subsubsection{Aims} 
Investigate the performance of the CD-Edgington estimator over a range of realistic scenarios and compare it to commonly used point and interval estimators.

\subsubsection{Data-Generating Mechanism}
The data-generating mechanism is adopted from \citet{heldpawelhofman2024}. We vary the number of studies $k \in$ \{3, 5, 10, 20, 50\}, the between-study heterogeneity determined by Higgins' $I^2$ $\in$ \{0\%, 30\%, 60\%, 90\%\}, the number of large studies $k_{\text{large}} \in$ \{0, 1, 2\} and whether the true effects $\ti$ are generated from a normal distribution or from a left-skewed skew-normal distribution, corresponding to model misspecification. Since there is no reason to assume that the direction of skewness affects the performance, we do not additionally consider a right-skewed effect distribution. Study sizes $n_i$ are set to 50 for normal studies and to 500 for large studies. We perform the simulation study in a full-factorial manner.

The true mean effect is set to $\mu = -0.3$. In absence of heterogeneity, this corresponds to the common effect; in the presence of heterogeneity, it represents the average effect. For consistency, we refer to it as the mean effect throughout, while implicitly acknowledging that its interpretation depends on the degree of heterogeneity. In each of $\niter$ iterations, we:

\begin{enumerate}
\item Simulate $k$ squared standard errors $\se(\hti)^2$ from a $\Chi$-distribution:
$$
\se(\hti)^2 \sim \frac{1}{(n_i - 1)n_i} \chi_{2 (n_i - 1)}^2.
$$
\item Compute $\taus$ as:
$$
\taus = \frac{1}{k} \sum_{i = 1}^k \frac{2}{n_i} \frac{I^2}{1 - I^2}.
$$
\item Simulate $k$ true effects $\ti$:
\begin{enumerate}
\item For a normal effect distribution, generate effects from a $\Nor(\mu, \taus)$.
\item For a skew-normal effect distribution, generate the effects from a $\SN(\xi, \omega, \alpha)$, parameterized as by \citet{Azzalini2013}. The parameters are obtained by moment-matching such that the mean equals $-0.3$ and the variance equals $\taus$: the skewness parameter is set to $\alpha = -4$, inducing a left-skewed distribution; the scale parameter is set to $\omega = \sqrt{\taus / (1 - 2 \delta^2 / \pi)}$, where $\delta = \alpha/\sqrt{1 + \alpha^2}$; the location parameter is set to $\xi = \mu - \omega  \delta \sqrt{2 / \pi}$. An example of a skew normal distribution is displayed in Figure~S2.

\end{enumerate}

\item Generate $k$ effect estimates $\hti$ on the standardized mean difference scale:
$$
\hti \sim \Nor(\ti, 2/n_i).
$$
\end{enumerate}

\subsubsection{Estimands and Other Targets}
The mean of the data-generating distribution is set to $\mu = -0.3$, which is the estimand for evaluating coverage and bias.

\subsubsection{Methods}
We compare the equi-tailed 95\% CD-Edgington confidence interval with the classical random-effects interval, the HKSJ interval \citep{Hartung2001, sidik2002simple} and with Edgington's method with additive heterogeneity adjustment. We evaluate the CD-Edgington point estimator together with the IVW point estimator, which is used in classical random-effects meta-analysis and in the HKSJ method, and the point estimator from Edgington's method with additive heterogeneity adjustment. For the CD-Edgington estimator, we use the Monte Carlo algorithm with 100{,}000 samples for the computation. The classical methods are accessed through the \textsf{R} \textsf{meta} package \citep{Rmeta}. Across all methods, the REML estimator is used for estimating between-study heterogeneity, recommended by \citet{Langan2018}

\subsubsection{Performance Measures}
The primary performance measure is the coverage of 95\% confidence intervals, estimated as the proportion of intervals overlapping the true mean effect. We perform $\niter = 4000$ iterations under each scenario, inducing a maximum Monte Carlo standard error (MCSE) (under a worst case scenario true coverage of 50\%) of:
$$
\text{MCSE}_{\widehat{\text{Cov}}} = \sqrt{\frac{\widehat{\text{Cov}} \ ( 1 - \widehat{\text{Cov}})}{4000}} \approx 0.008.
$$
To assess confidence validity \citep{morris2019using}, we examine interval coverage and width jointly. Further, we investigate the skewness of 95\% confidence intervals by examining the correlation and agreement between the skewness of intervals and of effect estimates $\hti$ and true effects $\ti$, respectively. Examining the skewness of $\hti$ and $\ti$ provides information on how asymmetry in confidence intervals relates to directly observed effect estimate skewness, but also skewness of parameters which the effect estimates are proxys for, thereby indirectly quantifying the distortion due to sampling noise. The skewness of $\hti$ is computed as Fisher's weighted skewness coefficient \citep{ferschl1980deskriptive}:
$$
\gamma = \frac{\left(\sum_{i=1}^{k} \frac{1}{\hsi} \left(\hti - \hmuivw^{\text{(fixed)}}\right)^3\right) \sqrt{\sum_{i=1}^{k} \frac{1}{\hsi}}}{ \left( \sum_{i=1}^{k} \frac{1}{\hsi} \left(\hti - \hmuivw^{\text{(fixed)}}\right)^2 \right)^{3/2}},
$$
while the skewness of $\ti$ is computed using Fisher's (unweighted) skewness coefficient, which is obtained from the above by setting all $\hsi$ to one. The correlation between $\beta$ and $\gamma$ is computed using Pearson's correlation coefficient, and Cohens kappa is used to quantify sign agreement. The classical random-effects and HKSJ intervals are always symmetric, and hence correlation and agreement cannot be estimated. With respect to the point estimator we evaluate bias and mean squared error (MSE).

\subsubsection{Computational Details}
The simulation study is programmed in the \textsf{R} programming language \citep{RRR} and conducted in \textsf{R} version 4.5.0 (2025-04-11) on a remote Debian GNU/Linux server (platform: x86\_64-pc-linux-gnu). Random number generator streams are employed to ensure reproducibility in parallel execution. The code, results and detailed information on the computational environment are publicly available on Github (\url{https://github.com/davidkronthaler-dk/sim-edgemeta.git}).

\subsection{Results}

We observed no non-convergences in the simulation study.

\subsubsection{95\% Confidence Intervals}
Figure~\ref{fig:cicovernor} displays the coverage of 95\% confidence intervals under normally distributed effects. Average interval widths are presented in Figure~\ref{fig:ciwnor}. The CD-Edgington confidence interval tends to exhibit coverage exceeding the nominal level under no heterogeneity, approaching nominal level as the number of studies increases. The random-effects interval exhibits a similar trend but typically remains closer to nominal coverage. The HKSJ method and Edgington's method with additive heterogeneity typically achieve nominal coverage under no heterogeneity, with Edgington's method exceeding nominal coverage in scenarios with one large and three to five studies. Despite its conservatism, the CD-Edgington interval only marginally differs in width compared to the HKSJ interval, being narrower under no large studies and wider under one or two large studies. Edgington's method with additive heterogeneity and the random-effects interval are typically narrower when three to five studies are included, likely due to not accounting for heterogeneity estimation uncertainty. For scenarios with more than five studies, interval widths are very similar under no heterogeneity.

Under heterogeneity, the CD-Edgington interval typically attains nominal coverage, with slight overcoverage for Higgins’ $I^2$ of 30\% and three studies and slight undercoverage for $I^2$ of 90\% and three to five studies. The HKSJ interval attains nominal coverage in all scenarios without large studies. With one or two large studies, coverage is generally too low for three to ten studies, except under $I^2$ of 90\%, where undercoverage occurs only with three studies.

Edgington's method with additive heterogeneity adjustment typically approaches nominal coverage under heterogeneity provided that at least ten studies are included. For fewer studies, coverage may be as low as 85\%. In scenarios with large studies and heterogeneity it typically outperforms the random-effects interval but yields consistently lower coverage than the CD-Edgington interval, and typically also lower coverage than the HKSJ method. The random-effects interval exhibits substantial undercoverage with ten or fewer studies under heterogeneity.

The CD-Edgington interval is typically narrower than the HKSJ interval, particularly evident under large heterogeneity, except in scenarios with three studies, one or two large studies and Higgins' $I^2$ of 0\% or 30\%. Both methods produce confidence intervals that are typically wider than the approaches not accounting for uncertainty in heterogeneity estimation, with differences diminishing as the number of studies increases.

Figure~S5 displays the Pearson correlation between the skewness of confidence intervals and effect estimates for normally distributed effects. Both approaches based on Edgington's method reflect the skewness of effect estimates effectively, with correlations never falling below 0.5 and sometimes approaching one. Correlations generally decrease as the number of studies increases. Under three or 50 studies, Edgington's method with additive heterogeneity typically exhibits larger correlations, while the CD-Edington interval does so for five to 20 studies. Similar trends are observed for Cohen's kappa assessing sign agreement (Figure~S7). Confidence intervals also capture the skewness of true effects reasonably well, though correlations and Cohen’s kappa are generally lower than for effect estimates, reflecting noise from sampling variability around true effects (Figures~S9 and~S11).

For true effects distributed according to a skew-normal distribution, Figures~S3,~S4,~S6,~S8,~S10 and~S12 display corresponding results. Coverages are very similar across effect distributions, with only slight decreases for effects distributed according to a skew-normal distribution, mainly observed under Higgins' $I^2$ of 90\%. Confidence interval widths and skewness results, depending only on the effect estimates and true effects, are virtually identical under both effect distributions.
{
\begin{landscape}
\begin{figure}
\centering
\begin{knitrout}
\definecolor{shadecolor}{rgb}{0.969, 0.969, 0.969}\color{fgcolor}

{\centering \includegraphics[width=1\linewidth]{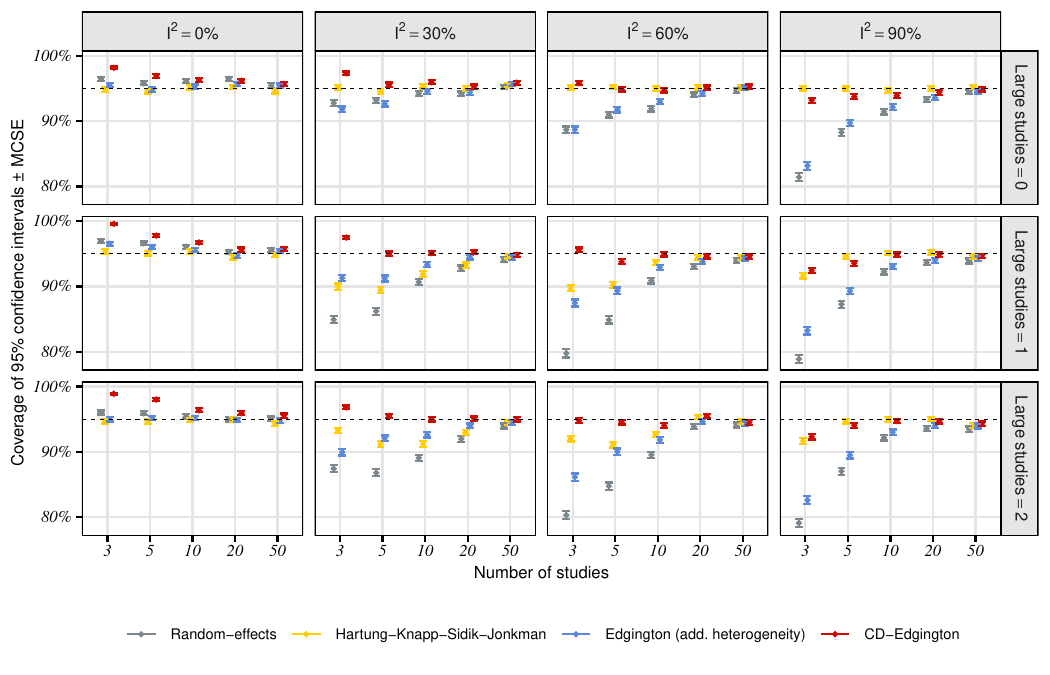} 

}

\end{knitrout}
\caption{Coverage of 95\% confidence intervals for the mean effect, for true effects following a normal distribution. Error bars represent Monte Carlo standard errors (MCSE).}
\label{fig:cicovernor}
\end{figure}
\end{landscape}

\begin{landscape}
\begin{figure}
\centering
\begin{knitrout}
\definecolor{shadecolor}{rgb}{0.969, 0.969, 0.969}\color{fgcolor}

{\centering \includegraphics[width=1\linewidth]{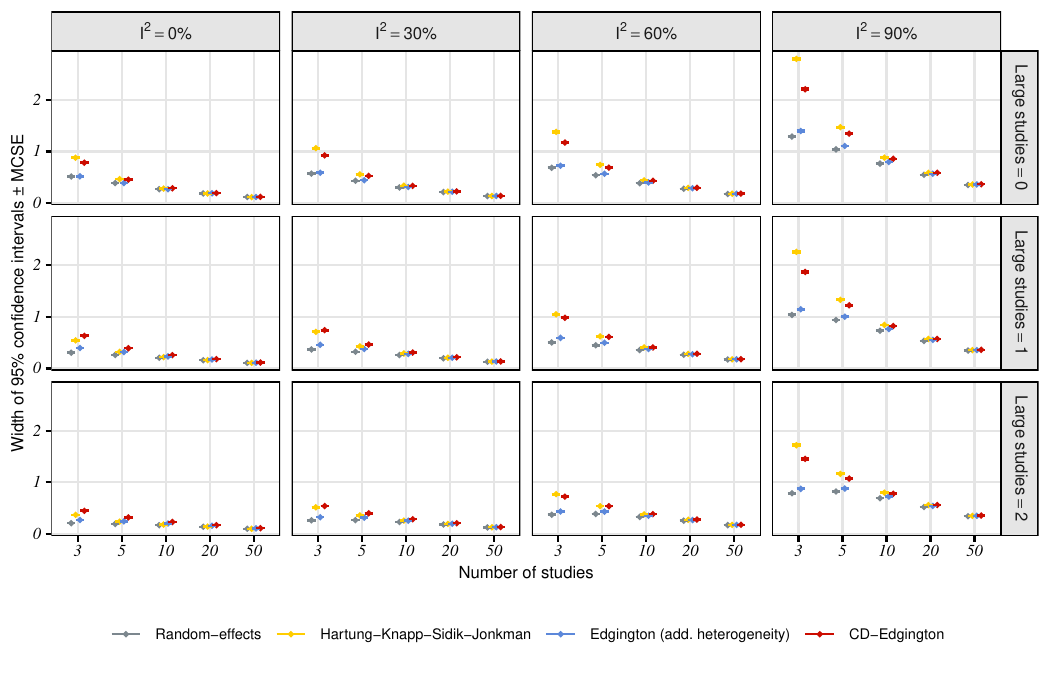} 

}

\end{knitrout}
\caption{Width of 95\% confidence intervals for the mean effect, for true effects distributed according to a normal distribution. Error bars represent Monte Carlo standard errors (MCSE).}
\label{fig:ciwnor}
\end{figure}
\end{landscape}
}

\subsubsection{Point Estimation}
The average bias of point estimators under normally distributed effects is presented in Figure~\ref{fig:biasnor}. All estimators are approximately unbiased for the true mean effect. Fewer studies and larger heterogeneity increase variability, while increasing the number of studies generally reduces bias. The strongest average bias observed occurs under Higgins' $I^2 of 90$\%, reaching $-0.0097$ for the random-effects estimator. The corresponding MSEs are comparable across methods and decrease as the number of studies increases and increase with larger heterogeneity (Figure~S14) . When true effects follow a skew-normal distribution, the bias of methods based on Edgington's approach systematically deviates from zero under scenarios with Higgins' $I^2$ of 60\% and 90\% (Figure~S13). Notably, the bias increases in the number of studies when $I^2$ is 90\%. Maximum average bias of 0.037 is observed for Edgington's method with additive heterogeneity adjustment under $I^2$ of 90\% and 50 studies. The average bias is consistently positive in these scenarios, reflecting the left-skewness of the skew-normal distribution. If the skew-normal distribution were right-skewed, we would expect the bias to be negative instead. In contrast, the random-effects estimator remains approximately unbiased across all scenarios. MSE trends resemble those observed under the normal effect distribution, with slightly higher MSEs under a skew-normal effect distribution (Figure~S15). 

\afterpage{
\begin{landscape}
\begin{figure}
\centering
\begin{knitrout}
\definecolor{shadecolor}{rgb}{0.969, 0.969, 0.969}\color{fgcolor}

{\centering \includegraphics[width=1\linewidth]{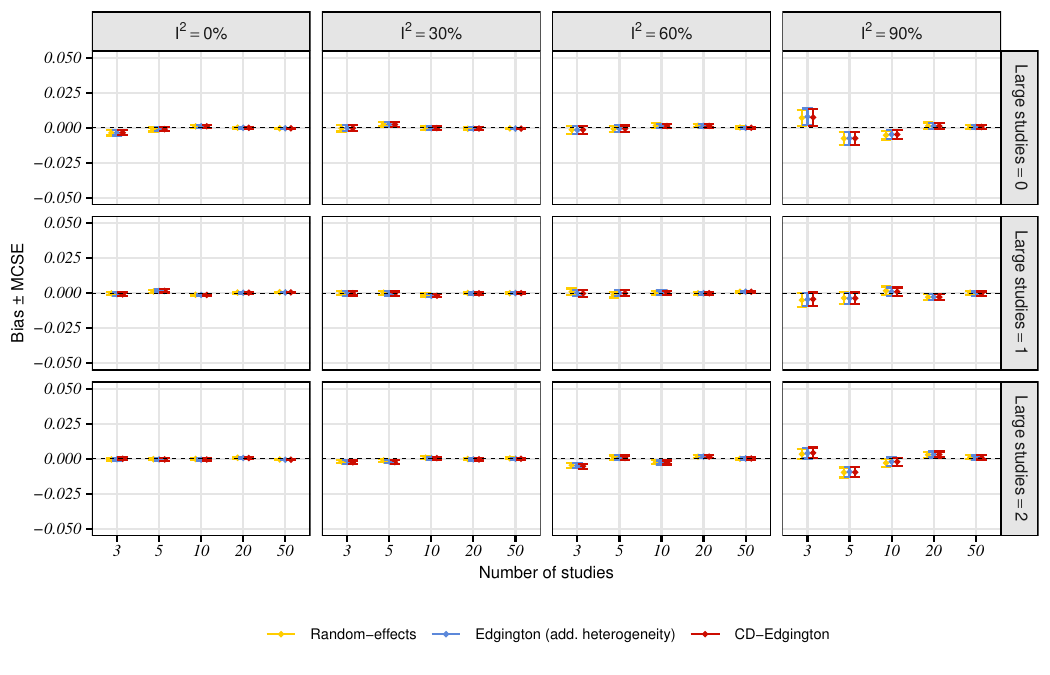} 

}

\end{knitrout}
\caption{Bias for the mean effect, for true effects distributed according to a normal distribution. Error bars represent Monte Carlo standard errors (MCSE).}
\label{fig:biasnor}
\end{figure}
\end{landscape}
}

\subsubsection{Summary of Simulation Results}
The simulation results suggest that the CD-Edgington estimator remains unbiased under correct model assumptions, and its 95\% confidence interval approaches nominal coverage under most scenarios including more than three studies and heterogeneity. Under no heterogeneity or for only three studies, it typically overcovers,  but is often narrower than the HKSJ interval. Under model misspecification, confidence interval coverage drops only slightly, while the point estimator exhibits small bias when heterogeneity is large. We replicated the simulation results of \citet{heldpawelhofman2024}: The 95\% confidence interval from Edgington's method with additive heterogeneity tends to undercover when the number of studies is small, but approaches nominal level with ten or more studies. However, the intervals are generally narrower than those from the HKSJ method or the CD-Edgington estimator.

\section{Discussion}
We proposed estimation of the average effect in random-effects meta-analysis using an extended version of Edgington's method. Our approach involves integration of Edgington's confidence distribution over a confidence distribution of the nuisance heterogeneity parameter to account for estimation uncertainty. Such marginalization generally produces an approximate confidence distribution, for which correct frequentist coverage is not guaranteed without further validation through simulations \citep{Schweder2016}. Our simulation results suggest that this approximation effectively incorporates uncertainty with respect to the heterogeneity parameter into the estimation of the average effect, yielding confidence intervals with coverage close to nominal level under five or more studies and heterogeneity, while typically outperforming Edgington's method without such estimation uncertainy adjustment.


We find that the HKSJ approach typically achieves nominal coverage when study sample sizes are equal, though this is rarely the case in practice. As already noted by \citet{IntHout2014}, with unequal study sizes, the HKSJ method may exhibit undercoverage with 20 or fewer studies for $I^2= 30$\%, 10 or fewer studies for $I^2 = 60$\%, and three studies for $I^2 = 90$\%. Despite this, the Cochrane Handbook recommends the HKSJ method when heterogeneity is estimated greater than zero and the number of studies exceeds two \citep{deeks2024chapter10}. This may require clarification regarding the influence of study sample sizes. Additionally, with only three studies, HKSJ intervals are typically wide due to the heavy tails of the $t$-distribution with one degree of freedom. In contrast, CD–Edgington intervals are less sensitive to study sample sizes, with coverage typically closer to nominal under heterogeneity.

Our method is subject to limitations: While computation can be performed using computationally efficient deterministic GAQ integration, simulations suggest that the produced intervals are too wide in scenarios with few studies. Alternatively, we proposed a Monte Carlo algorithm. A drawback is that fully stable results require a substantial number of samples, which may not be computationally feasible \citep{Nagashima2018}. Instead, the choice of samples could be informed by approaches such as that of \citet{gelman1992inference}. 

Further, simulations suggest that CD-Edgington intervals overcover under no heterogeneity or when only three studies are available. Applying a frequentist random-effects meta-analysis with so few studies has been previously considered unreliable, as between-study heterogeneity cannot be estimated precisely \citep{Lilienthal2023}. For this reason, when only few studies are available, a fixed-effect analysis is sometimes recommended; alternatively, researchers may compare multiple random-effects methods or rely on a qualitative synthesis \citep{bender2018methods}. Well-informed Bayesian or empirical Bayes approaches can also yield more reliable results than purely frequentist methods \citep{rover2023summarizing, Lilienthal2023}.

Our simulations were limited to effect estimates generated on the standardized mean difference scale. Extending these to logarithmized odds, risk, or hazard ratios, which are commonly used to accommodate binary or survival outcomes \citep{borenstein2021introduction}, would provide a more conclusive assessment of the methods’ performance. Future research could also explore accommodating the proposed heterogeneity uncertainty adjustment to alternative $p$-value combination methods beyond Edgington’s approach. While a set of methods has been considered unreliable due to lack of orientation invariance \citep{heldpawelhofman2024}, it would be interesting to apply the approach to classical meta-analysis, that is, a weighted Stouffer $p$-value combination \citep{Senn2021}. A related approach using the confidence distribution of the standard $\Q$ statistic has been explored by \citet{Nagashima2018} for prediction intervals, but not for estimation. Future work could also investigate extending the methodology to meta-regression and cumulative meta-analysis. Further, \citet{held2024assessment} proposed a weighted version of Edgington’s method for replication studies. While the original method already incorporates weighting via the slopes of study-specific $p$-value functions, additional weights can be introduced to downweight studies at risk of bias. We plan to investigate this extension in the future.

In this work, we proposed methods to incorporate uncertainty in heterogeneity estimation when estimating the average effect in random-effects meta-analysis using the Edgington combined $p$-value function. However, we have not yet addressed another central aim of meta-analysis: prediction of future study effects \citep{Higgins2008}. According to \citet[][p.~38]{Viechtbauer2006}, \enquote{quantifying the amount of heterogeneity and exploring its sources are among the most important aspects of systematic reviews}. Recent literature emphasizes predictive distributions and intervals as key tools for this purpose. Accordingly, part two of this series focuses on prediction.

%
%
%
%

\newpage
\bibliographystyle{apalike}
\bibliography{paper-ref}

\newpage
\renewcommand{\thesection}{S\arabic{section}}
\renewcommand{\thetable}{S\arabic{table}}
\renewcommand{\thefigure}{S\arabic{figure}}

\setcounter{section}{0}
\setcounter{table}{0}
\setcounter{figure}{0}

\clearpage
\renewcommand{\thepage}{S\arabic{page}} 
\setcounter{page}{1}

\begin{center}
    \LARGE\bfseries Supplementary Material\\[1em]
\end{center}

\vspace{2em}
\section{The Edgington Combined $p$-Value Function as a Confidence Distribution}\label{app:edgington}

Here we formally justify interpreting the Edgington combined $p$-value function as a confidence distribution according to Definition~1 in the main text. Specifically, the Edgington combined $p$-value function of the parameter $\mu$ (the average effect under the random-effects model), denoted by $p_{\text{E}}(\mu)$, corresponds to evaluating the cumulative distribution function (CDF) of the Irwin--Hall (IH) distribution with parameter $k$ at the sum of one-sided $p$-values $p_i(\mu)$, $i \in \{1,\ldots,k\}$, from $k$ individual studies \citep{edgington1972additive}:
$$
p_{\text{E}}(\mu) = F_{\text{IH},k}\left(\sum_{i = 1}^k p_i(\mu)\right).
$$
The $k$ study-specific $p$-value functions are derived as
$$
Z_i(\mu) = \frac{\hti - \mu}{\sqrt{\htaus + \hsi}} \sim \Nor(0, 1), \quad p_i(\mu) = 1 - \Phi \left(Z_i(\mu)\right),
$$
where $\Phi(\cdot)$ denotes the CDF of the standard normal distribution, $\hti$ denotes the study estimate, $\hsi$ its squared standard error, and $\htaus$ the estimated between-study variance. Therefore, $\forall \mu \in \mathbb{R}, \forall \hat{\theta}_1, \ldots, \hat{\theta}_k \in \mathbb{R}^k, \forall \hat{\sigma}^2_1, \ldots, \hat{\sigma}^2_k \in \mathbb{R}^k_+, \ p_{\text{E}}(\mu): \mathbb{R} \mapsto [0, 1]$. Further, the Edgington combined $p$-value function is a monotonically increasing and right-continuous function with respect to $\mu$, since:

\begin{enumerate}
  \item The individual $p$-value functions $p_i(\mu)$ are monotonically increasing in $\mu$;
  \item Consequently, the sum of $p$-values $\sum_{i = 1}^k p_i(\mu)$ is monotonically increasing in $\mu$;
  \item The CDF of the IH distribution maps this sum to the unit interval without affecting monotonicity and ensures right-continuity.
\end{enumerate}
Given monotonicity of the sum of the $p$-values and the IH distribution, the following limit conditions hold:
$$
\lim_{\mu \to -\infty} p_{\text{E}}(\mu) = 0 \quad \text{and} \quad \lim_{\mu \to \infty} p_{\text{E}}(\mu) = 1.
$$
Hence, the Edgington combined $p$-value function is a CDF of the parameter $\mu$. Moreover, under mild conditions, the Edgington combined $p$-value function evaluated at the true parameter value, $\mu = \mu_0$, converges in distribution to a standard uniform random variable.

\begin{assumption} 
The estimators $\hti, i \in \{1,\dots,k\}, $ are independent and normally distributed under the true value $\mu = \mu_0$: $\hti \sim \Nor(\mu_0, \htaus + \hsi)$.
\end{assumption}
\begin{assumption}
The squared standard errors $\hsi, i \in \{1,\dots,k\}$, are mutually independent and independent of the corresponding estimators $\hti$.
\end{assumption}

\begin{step}
Uniformity of individual $p$-values: Each one-sided $p$-value under $\mu = \mu_0$ satisfies $p_i(\mu_0) = 1 - \Phi \left(Z_i(\mu_0)\right) \sim \Uni[0, 1]$, implied by the probability integral transform (PIT).
\end{step}

\begin{step}
Approximate independence of the $p$-values: The $p$-values are approximately independent, assuming that any dependence introduced by shared $\htaus$ is negligible:
$$
p_i(\mu_0) \sima \text{i.i.d. } \Uni[0, 1].
$$
\end{step}

\begin{step}
Distribution of the sum of $p$-values: The IH distribution with parameter $k$ is the distribution of the sum of $k$ independent standard uniform random variables. Therefore, for $k$ approximately independent $p$-values at $\mu = \mu_0$, we have:
$$
\sum_{i=1}^k p_i(\mu_0) \sima \text{IH}(k).
$$
\end{step}

\begin{step}
Application of the IH distribution: Hence by PIT:
$$
p_{\text{E}}(\mu_0) = F_{\text{IH},k}\left( \sum_{i=1}^k p_i(\mu_0) \right) \sima \Uni[0, 1].
$$
\end{step}

\begin{step}
Central limit theorem (CLT) approximation for $k \ge 12$: When using a CLT argument to approximate the IH distribution for $k \ge 12$, the steps to derive the desired properties for this approximation are analogous to the ones described above for the exact IH distribution.
\end{step}
\noindent Hence, it follows that the Edgington combined $p$-value function defines the CDF of an \emph{approximate confidence distribution} of $\mu$.

\section{Confidence Density of the Between-Study Heterogeneity}\label{app:qderivative}

In the Methods section of the main text we discuss the confidence distribution of the heterogeneity parameter $\taus$, induced by the generalized heterogeneity statistic $\Q(\taus)$ \citep{Viechtbauer2006}. The confidence density of $\taus$ can be obtained by change of variables: Since $\Q(\taus)$ is monotonically decreasing in $\taus$, its derivative is well-defined, and $\taus = \Q^{-1}(\Q(\taus))$, the confidence density of $\taus$ is
$$
c(\taus) = f_{\chi_{k-1}^2}(\Q(\taus)) \left| { \frac{\d \Q(\taus)}{\d \taus}} \right|,
$$
where $f_{\chi_{k-1}^2}(\cdot)$ denotes the density of a $\Chi$-distribution with $k-1$ degrees of freedom. Here we present the computation of the derivative of $\Q(\taus)$ with respect to $\taus$. We denote the weight of study $i$, $i \in \{1,\ldots,k\}$, as $\wit = 1/(\taus + \hsi)$. Hence, by product and quotient rule:
$$
\frac{\d\Q(\taus)}{\d\taus} = \sum_{i=1}^k \left(\frac{\d \wit}{\d \taus} \left( \hti - \hmuivw(\taus) \right)^2 - 2 \wit \left( \hti - \hmuivw(\taus) \right) \frac{\d \hmuivw(\taus)}{\d \taus} \right),
$$
where $\hmuivw$ denotes the classical inverse-variance weights (IVW) estimator for random-effects meta-analysis. The derivative of the weights is:
$$
\frac{\d \wit}{\d \taus} = \frac{-1}{(\si + \taus)^2}.
$$
Now we denote
$$
\A(\taus) = \sum_{i=1}^k \wit \hti, \quad \B(\taus) = \sum_{i=1}^k \wit, \quad \hmuivw(\taus) = \frac{\A(\taus)}{\B(\taus)}.
$$
Then the derivative of the IVW estimator with respect to $\taus$ is
$$
\frac{\d \hmuivw(\taus)}{\d \taus} = \frac{\B(\taus) \frac{\d \A(\taus)}{\d \taus} - \A(\taus) \frac{\d \B(\taus)}{\d \taus}}{\B(\taus)^2},
$$
with
$$
\frac{\d \A(\taus)}{\d\taus} = \sum_{i=1}^k \frac{\d \wit}{\d \taus} \hti, \quad \frac{\d \B(\taus)}{\d\taus} = \sum_{i=1}^k \frac{\d \wit}{\d \taus}.
$$
Hence, we obtain an expression for the analytic derivative of the generalized heterogeneity statistic with respect to the heterogeneity parameter $\taus$,
$$
\frac{\d \Q(\taus)}{\d\taus} = \sum_{i=1}^k \left[- \frac{1}{(\si + \taus)^2} \left( \hti - \hmuivw(\taus) \right)^2 + \frac{2}{\si + \taus} \left( \hti - \hmuivw(\taus) \right) \frac{\d \hmuivw(\taus)}{\d\taus}
\right],
$$
which in turn enables the computation of the confidence density of $\taus$. 

\begin{figure}
\centering
\begin{knitrout}
\definecolor{shadecolor}{rgb}{0.969, 0.969, 0.969}\color{fgcolor}

{\centering \includegraphics[width=0.9\linewidth]{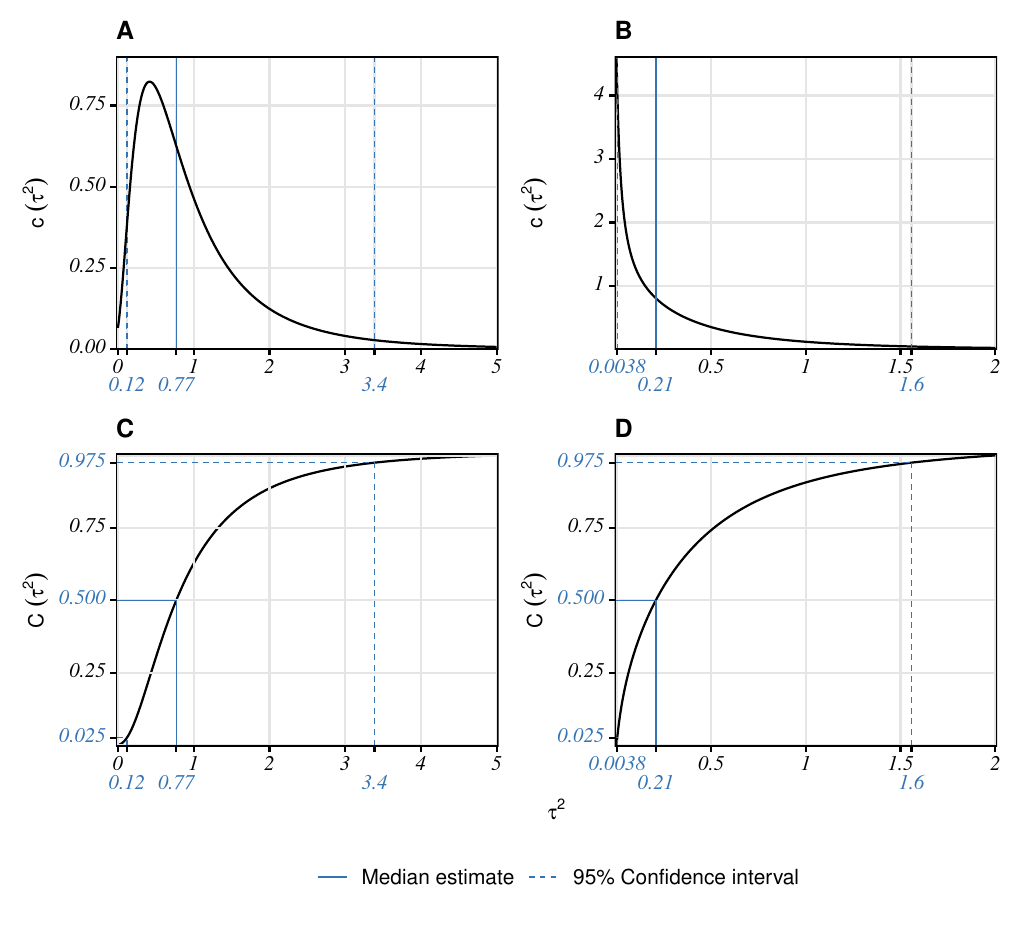} 

}

\end{knitrout}
\caption{Confidence densities and confidence distribution functions of the between-study heterogeneity parameter $\taus$, based on the generalized heterogeneity statistic. Panels (A, C) correspond to nine reported mean differences on \textit{Serenoa repens} treatment for urinary tract symptoms \citep{Franco2023}, and panels (B, D) correspond to seven reported log odds ratios quantifying the association between corticosteroids and mortality in hospitalized COVID-19 patients \citep{who2020corticosteroids}.}

\label{fig:exampleftau2}
\end{figure}

To illustrate the applicability of the derived confidence density, Figure~\ref{fig:exampleftau2} displays the confidence densities and confidence distribution functions of $\taus$, based on two meta-analyses. The first is based on nine reported mean differences investigating the effect of \textit{Serenoa repens} treatment on lower urinary tract symptoms \citep{Franco2023}, yielding $\remltaus$ of 0.85 (95\% confidence interval from 0.11 to 3.96; Higgins' $I^2$ = 67.36\%), estimated using the restricted maximum likelihood (REML) approach. The second example uses seven reported log odds ratios quantifying the association between corticosteroids and mortality in hospitalized COVID-19 patients \citep{who2020corticosteroids}, with an estimated $\remltaus < 0.0001$ (95\% confidence interval from 0.00 to 2.13; Higgins' $I^2$ = 14.01\%). 

In the first example, where heterogeneity is substantial, the confidence distribution is broad and peaks away from zero. In contrast, the second example with small heterogeneity produces a density sharply peaked at zero. The median estimates for $\taus$ obtained from the confidence distribution approach presented here are 0.77 (95\% confidence interval from 0.12 to 3.39) and 0.21 (95\% confidence interval from 0.00 to 1.56), respectively. While these summaries are provided for comparison with REML estimates, we emphasize that this approach is designed not to reduce the confidence distribution of $\taus$ to a scalar value or interval, but rather to represent uncertainty in the form of a full confidence distribution.

\section{CD-Edgington: Monte Carlo and Global Adaptive Quadrature}

For the computation of the marginalized confidence distribution of the parameter $\mu$,
$$
c(\mu) = \int c(\mu \mid \taus) \ c(\taus) \ \d \taus,
$$
\noindent we proposed a Monte Carlo algorithm and discussed deterministic global adaptive quadrature integration. Here, we present the results of a pilot simulation with 1000 iterations comparing the two approaches. We used the design of the simulation study presented in the main text and varied the number of studies $k \in \{3,5,10,20,50\}$ and between-study heterogeneity quantified by Higgins' $I^2 \in \{0\%, 30\%, 60\%, 90\%\}$ for normally distributed true effects and no large studies. Table~\ref{tab:MCvsGAQ} displays mean differences in point estimates and 95\% confidence interval limits between the two integration approaches. Table~\ref{tab:MCvsGAQ_bc} shows bias of point estimators and coverage of 95\% confidence intervals for both methods.

\begin{table}[ht]
\centering
\caption{Mean differences with Monte Carlo standard errors in point estimates and 95\% confidence interval limits between Monte Carlo sampling and global adaptive quadrature integretation approaches.} 
\label{tab:MCvsGAQ}
\begin{tabular}{lllll}
  Estimate (MC - GAQ) &  &  &  &  \\ 
  Studies ($k$) & $I^2$ = 0\% & 30\% & 60\% & 90\% \\ 
  3 & -0.000 [0.000] & -0.000 [0.000] & -0.000 [0.000] & -0.000 [0.000] \\ 
  5 & 0.000 [0.000] & 0.000 [0.000] & -0.000 [0.000] & -0.000 [0.000] \\ 
  10 & -0.000 [0.000] & -0.000 [0.000] & 0.000 [0.000] & 0.000 [0.000] \\ 
  20 & -0.000 [0.000] & -0.000 [0.000] & 0.000 [0.000] & 0.000 [0.000] \\ 
  50 & 0.000 [0.000] & -0.000 [0.000] & 0.000 [0.000] & -0.000 [0.000] \\ 
  Lower 95\% CI (MC - GAQ) &  &  &  &  \\ 
  Studies ($k$) & $I^2$ = 0\% & 30\% & 60\% & 90\% \\ 
  3 & 0.069 [0.003] & 0.056 [0.003] & 0.037 [0.003] & -0.088 [0.005] \\ 
  5 & 0.030 [0.001] & 0.029 [0.001] & 0.025 [0.001] & 0.010 [0.000] \\ 
  10 & 0.009 [0.000] & 0.009 [0.000] & 0.008 [0.000] & 0.004 [0.000] \\ 
  20 & 0.004 [0.000] & 0.004 [0.000] & 0.004 [0.000] & 0.002 [0.000] \\ 
  50 & 0.002 [0.000] & 0.002 [0.000] & 0.002 [0.000] & 0.002 [0.000] \\ 
  Upper 95\% CI (MC - GAQ) &  &  &  &  \\ 
  Studies ($k$) & $I^2$ = 0\% & 30\% & 60\% & 90\% \\ 
  3 & -0.069 [0.003] & -0.056 [0.003] & -0.037 [0.003] & 0.088 [0.005] \\ 
  5 & -0.030 [0.001] & -0.029 [0.001] & -0.025 [0.001] & -0.010 [0.000] \\ 
  10 & -0.009 [0.000] & -0.009 [0.000] & -0.008 [0.000] & -0.004 [0.000] \\ 
  20 & -0.004 [0.000] & -0.004 [0.000] & -0.004 [0.000] & -0.002 [0.000] \\ 
  50 & -0.002 [0.000] & -0.002 [0.000] & -0.002 [0.000] & -0.002 [0.000] \\ 
    
\multicolumn{5}{l}{\footnotesize CI = confidence interval, GAQ = global adaptive quadrature, MC = Monte Carlo.} \\ 
\end{tabular}
\end{table}

\begin{table}[ht]
\centering
\caption{Bias of point estimators and coverage of 95\% confidenc intervals with Monte Carlo standard errors for Monte Carlo sampling and global adaptive quadrature integretation approaches.} 
\label{tab:MCvsGAQ_bc}
\begin{tabular}{lllll}
  MC: Bias &  &  &  &  \\ 
  Studies ($k$) & $I^2$ = 0\% & 30\% & 60\% & 90\% \\ 
  3 & 0.004 [0.004] & -0.001 [0.004] & -0.001 [0.005] & -0.007 [0.007] \\ 
  5 & 0.000 [0.003] & 0.001 [0.003] & 0.001 [0.003] & 0.003 [0.005] \\ 
  10 & -0.002 [0.002] & -0.002 [0.002] & -0.002 [0.002] & -0.003 [0.003] \\ 
  20 & 0.002 [0.001] & -0.001 [0.001] & 0.001 [0.001] & 0.002 [0.002] \\ 
  50 & 0.001 [0.001] & -0.000 [0.001] & 0.000 [0.001] & -0.000 [0.001] \\ 
  GAQ: Bias &  &  &  &  \\ 
  Studies ($k$) & $I^2$ = 0\% & 30\% & 60\% & 90\% \\ 
  3 & 0.004 [0.004] & -0.001 [0.004] & -0.000 [0.005] & -0.007 [0.007] \\ 
  5 & 0.000 [0.003] & 0.001 [0.003] & 0.001 [0.003] & 0.003 [0.005] \\ 
  10 & -0.001 [0.002] & -0.002 [0.002] & -0.002 [0.002] & -0.003 [0.003] \\ 
  20 & 0.002 [0.001] & -0.001 [0.001] & 0.001 [0.001] & 0.002 [0.002] \\ 
  50 & 0.001 [0.001] & -0.000 [0.001] & 0.000 [0.001] & -0.000 [0.001] \\ 
  MC: 95\% CI coverage &  &  &  &  \\ 
  Studies ($k$) & $I^2$ = 0\% & 30\% & 60\% & 90\% \\ 
  3 & 0.980 [0.004] & 0.981 [0.004] & 0.971 [0.005] & 0.960 [0.006] \\ 
  5 & 0.973 [0.005] & 0.965 [0.006] & 0.959 [0.006] & 0.940 [0.008] \\ 
  10 & 0.963 [0.006] & 0.962 [0.006] & 0.957 [0.006] & 0.960 [0.006] \\ 
  20 & 0.966 [0.006] & 0.961 [0.006] & 0.952 [0.007] & 0.951 [0.007] \\ 
  50 & 0.954 [0.007] & 0.959 [0.006] & 0.959 [0.006] & 0.959 [0.006] \\ 
  GAQ: 95\% CI coverage &  &  &  &  \\ 
  Studies ($k$) & $I^2$ = 0\% & 30\% & 60\% & 90\% \\ 
  3 & 0.999 [0.001] & 1.000 [0.000] & 0.993 [0.003] & 0.976 [0.005] \\ 
  5 & 0.997 [0.002] & 0.992 [0.003] & 0.987 [0.004] & 0.954 [0.007] \\ 
  10 & 0.972 [0.005] & 0.973 [0.005] & 0.973 [0.005] & 0.971 [0.005] \\ 
  20 & 0.972 [0.005] & 0.970 [0.005] & 0.964 [0.006] & 0.955 [0.007] \\ 
  50 & 0.960 [0.006] & 0.965 [0.006] & 0.966 [0.006] & 0.965 [0.006] \\ 
    
\multicolumn{5}{l}{\footnotesize CI = confidence interval, GAQ = global adaptive quadrature, MC = Monte Carlo.} \\ 
\end{tabular}
\end{table}

\section{Additional Simulation Results}\label{app:simresults}

\begin{figure}
\centering
\begin{knitrout}
\definecolor{shadecolor}{rgb}{0.969, 0.969, 0.969}\color{fgcolor}

{\centering \includegraphics[width=0.6\linewidth]{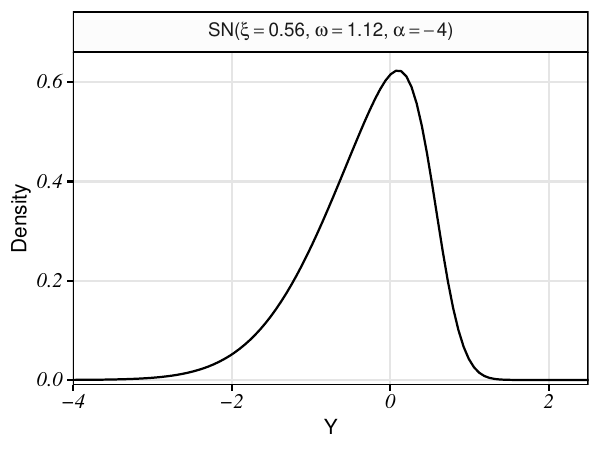} 

}

\end{knitrout}
\caption{Density of a skew-normal distribution with mean $-0.3$, variance $0.5$ and skewness parameter $\alpha = -4$.}
\label{fig:skewnormalexample}
\end{figure}

Table~\ref{tab:appendix_summary} provides an overview of simulation results presented in the Supplementary Material. An exemplary visualization of a skew-normal distribution parametrized as used in the simulation study is displayed in Figure~\ref{fig:skewnormalexample}.

\begin{table}[ht]
\centering
\caption{Simulation results and corresponding figure references presented in the Supplementary Material.}
\begin{tabularx}{\linewidth}{@{}X c c@{}}
{Performance measure} & {Effect distribution} & {Figure} \\[4pt]
\hline

Coverage of 95\% confidence intervals & Skew-normal & \ref{fig:cicoversn} \\[4pt]

Width of 95\% confidence intervals & Skew-normal & \ref{fig:ciwsn} \\[4pt] 

Pearson correlation between skewness of 95\% confidence intervals and skewness of effect estimates 
& Normal / Skew-normal 
& \ref{fig:ciskhesnor} / \ref{fig:ciskhessn} \\[4pt]

Cohen's kappa for sign agreement between skewness of 95\% confidence intervals and skewness of effect estimates
& Normal / Skew-normal & \ref{fig:cikappanor} / \ref{fig:cikappasn} \\[4pt]

Pearson correlation between skewness of 95\% confidence intervals and skewness of true effects 
& Normal / Skew-normal 
& \ref{fig:ciskesnor} / \ref{fig:ciskessn} \\[4pt]

Cohen's kappa for sign agreement between skewness of 95\% confidence intervals and skewness of true effects 
& Normal / Skew-normal 
& \ref{fig:cikappanores} / \ref{fig:cikappasnes} \\[4pt]

Bias of point estimators & Skew-normal & \ref{fig:biassn} \\[4pt]

Mean squared error (MSE) of point estimators & Normal / Skew-normal & \ref{fig:msenor} / \ref{fig:msesn} \\

\end{tabularx}
\label{tab:appendix_summary}
\end{table}

\begin{landscape}
\begin{figure}
\centering
\begin{knitrout}
\definecolor{shadecolor}{rgb}{0.969, 0.969, 0.969}\color{fgcolor}

{\centering \includegraphics[width=1\linewidth]{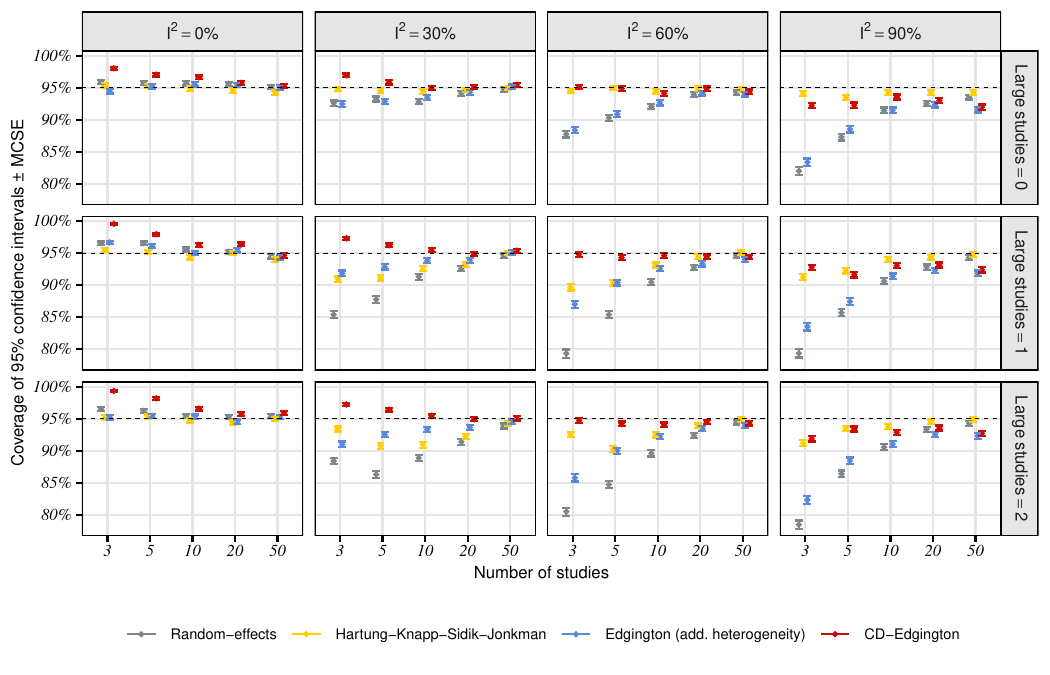} 

}

\end{knitrout}
\caption{Coverage of 95\% confidence intervals for the mean effect, for true effects following a left-skewed skew-normal distribution. Error bars represent Monte Carlo standard errors (MCSE).}
\label{fig:cicoversn}
\end{figure}
\end{landscape}

\begin{landscape}
\begin{figure}
\centering
\begin{knitrout}
\definecolor{shadecolor}{rgb}{0.969, 0.969, 0.969}\color{fgcolor}

{\centering \includegraphics[width=1\linewidth]{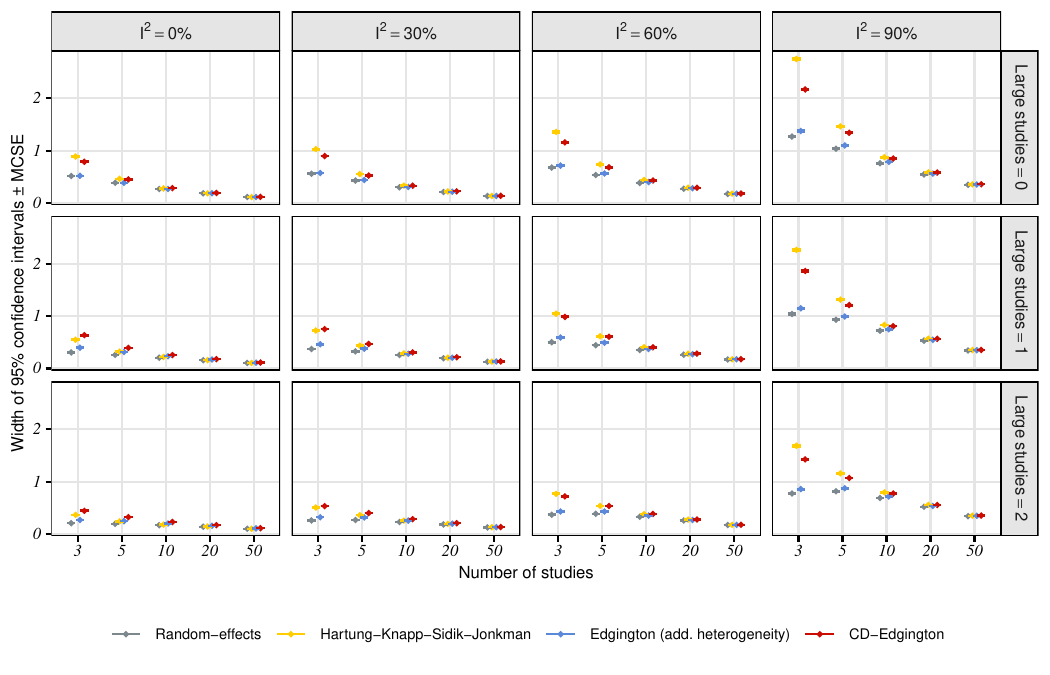} 

}

\end{knitrout}
\caption{Width of 95\% confidence intervals for the mean effect, for true effects distributed according to a left-skewed skew-normal distribution. Error bars represent Monte Carlo standard errors (MCSE).}
\label{fig:ciwsn}
\end{figure}
\end{landscape}

\begin{landscape}
\begin{figure}
\centering
\begin{knitrout}
\definecolor{shadecolor}{rgb}{0.969, 0.969, 0.969}\color{fgcolor}

{\centering \includegraphics[width=1\linewidth]{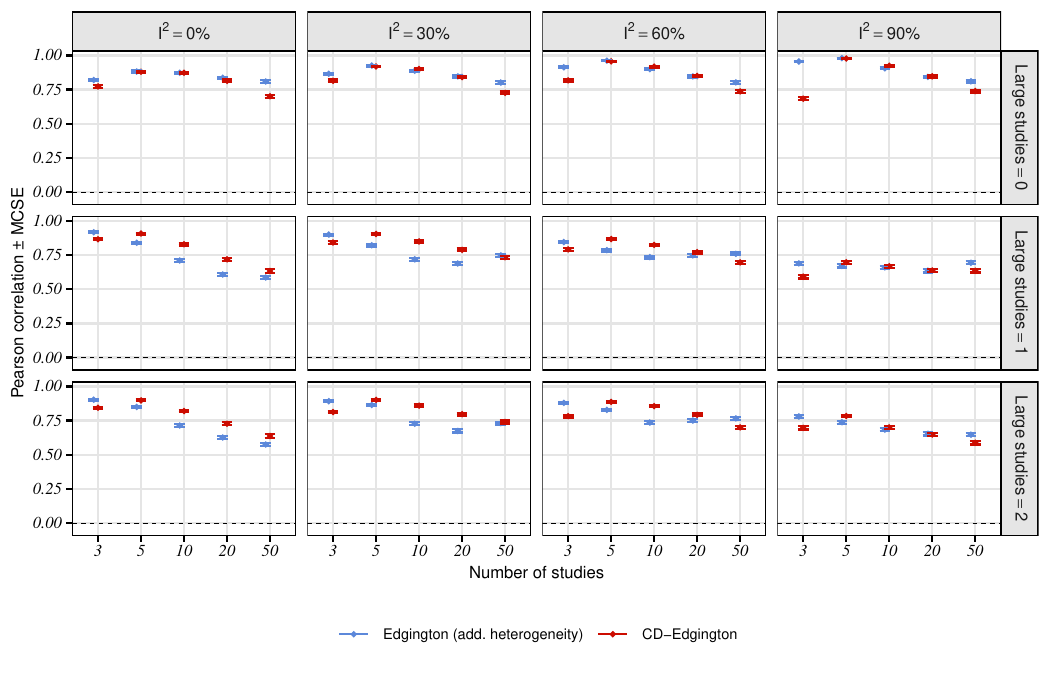} 

}

\end{knitrout}
\caption{Pearson correlation between the skewness of 95\% confidence intervals and the skewness of effect estimates, for true effects distributed according to a normal distribution. Error bars represent Monte Carlo standard errors (MCSE).}
\label{fig:ciskhesnor}
\end{figure}
\end{landscape}

\begin{landscape}
\begin{figure}
\centering
\begin{knitrout}
\definecolor{shadecolor}{rgb}{0.969, 0.969, 0.969}\color{fgcolor}

{\centering \includegraphics[width=1\linewidth]{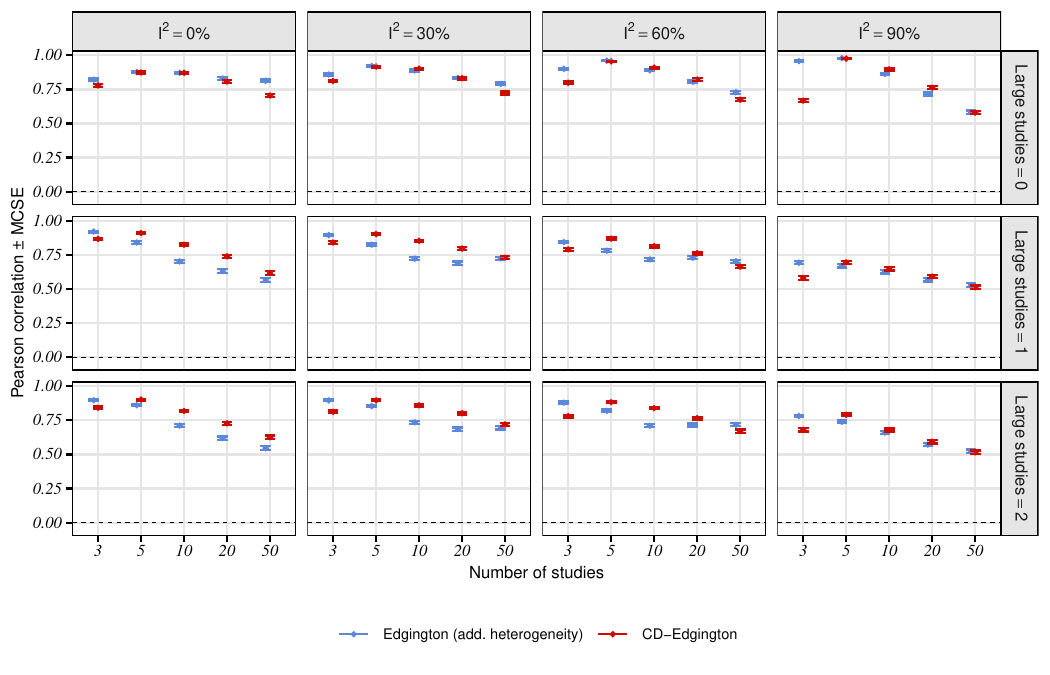} 

}

\end{knitrout}
\caption{Pearson correlation between the skewness of 95\% confidence intervals and the skewness of effect estimates, for true effects distributed according to a left-skewed skew-normal distribution. Error bars represent Monte Carlo standard errors (MCSE).}
\label{fig:ciskhessn}
\end{figure}
\end{landscape}

\begin{landscape}
\begin{figure}
\centering
\begin{knitrout}
\definecolor{shadecolor}{rgb}{0.969, 0.969, 0.969}\color{fgcolor}

{\centering \includegraphics[width=1\linewidth]{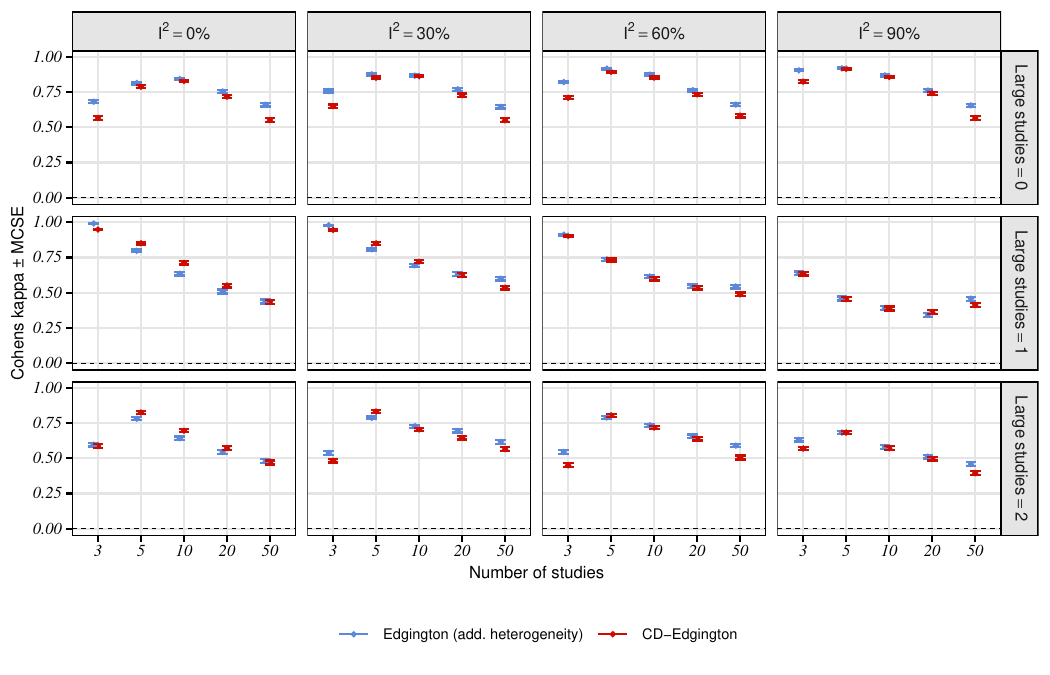} 

}

\end{knitrout}
\caption{Cohens kappa for sign agreement between the skewness of 95\% confidence intervals and the skewness of effect estimates, for true effects distributed according to a normal distribution. Error bars represent Monte Carlo standard errors (MCSE).}
\label{fig:cikappanor}
\end{figure}
\end{landscape}

\begin{landscape}
\begin{figure}
\centering
\begin{knitrout}
\definecolor{shadecolor}{rgb}{0.969, 0.969, 0.969}\color{fgcolor}

{\centering \includegraphics[width=1\linewidth]{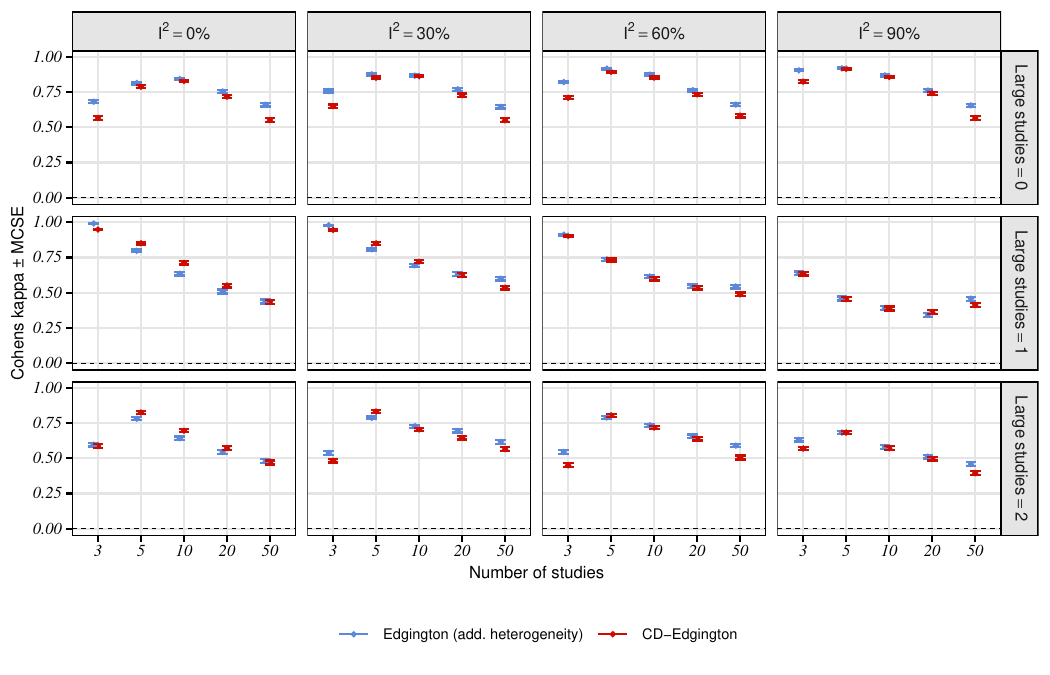} 

}

\end{knitrout}
\caption{Cohens kappa for sign agreement between the skewness of 95\% confidence intervals and the skewness of effect estimates, for true effects distributed according to a left-skewed skew-normal distribution. Error bars represent Monte Carlo standard errors (MCSE).}
\label{fig:cikappasn}
\end{figure}
\end{landscape}

\begin{landscape}
\begin{figure}
\centering
\begin{knitrout}
\definecolor{shadecolor}{rgb}{0.969, 0.969, 0.969}\color{fgcolor}

{\centering \includegraphics[width=1\linewidth]{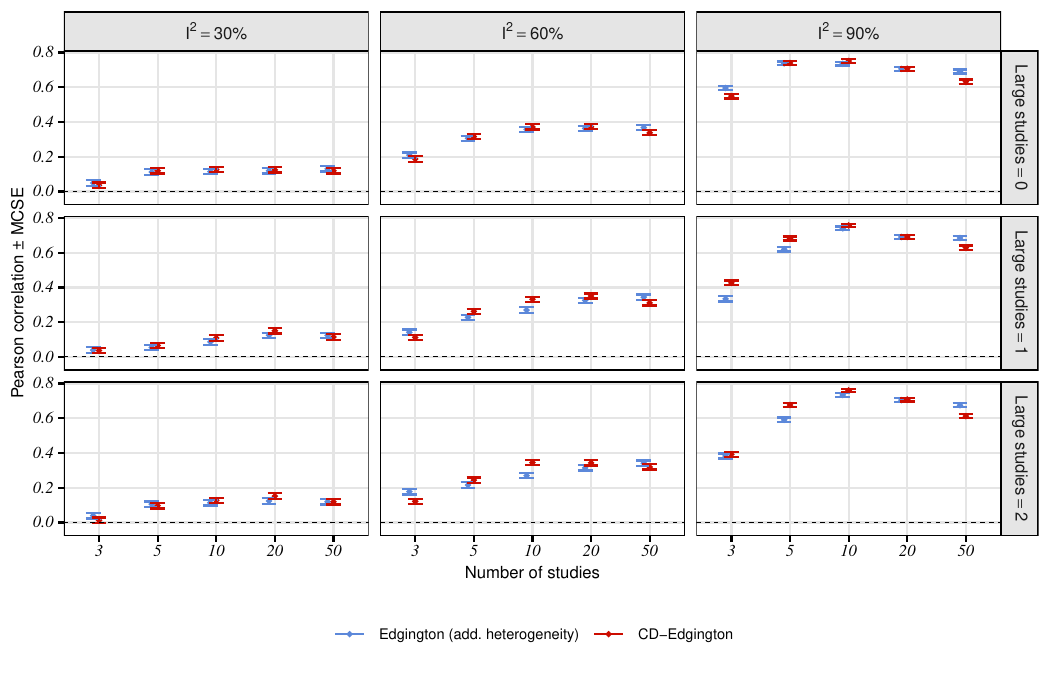} 

}

\end{knitrout}
\caption{Pearson correlation between the skewness of 95\% confidence intervals and the skewness of normal true effects. Scenarios with no true heterogeneity are omitted, since then all true effects equal the true mean effect. Error bars represent Monte Carlo standard errors (MCSE).}
\label{fig:ciskesnor}
\end{figure}
\end{landscape}

\begin{landscape}
\begin{figure}
\centering
\begin{knitrout}
\definecolor{shadecolor}{rgb}{0.969, 0.969, 0.969}\color{fgcolor}

{\centering \includegraphics[width=1\linewidth]{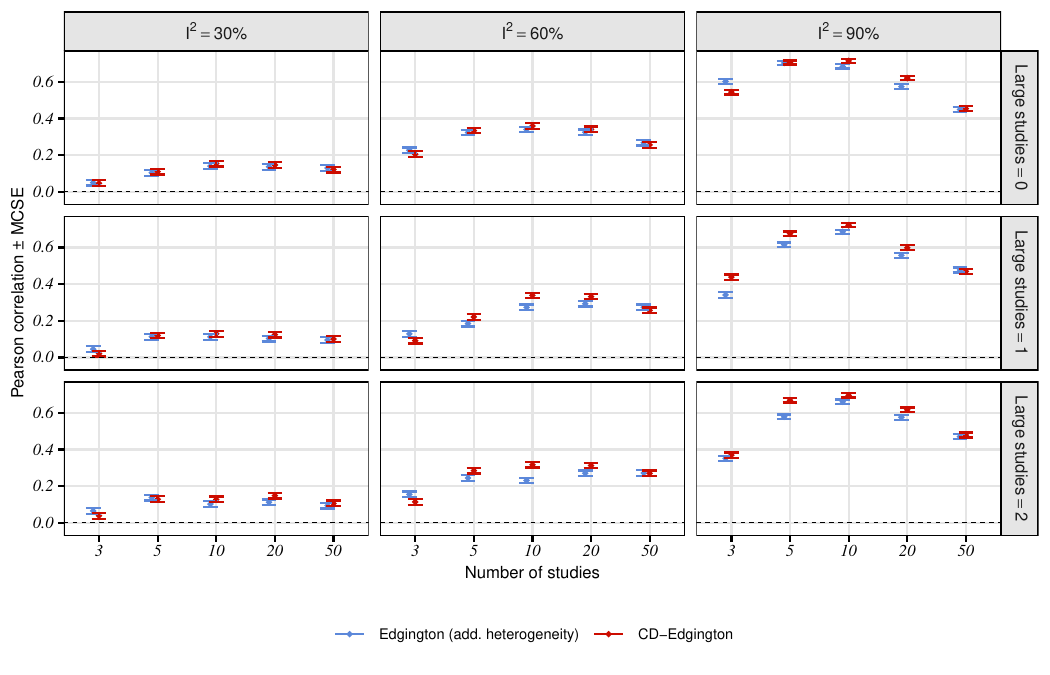} 

}

\end{knitrout}
\caption{Pearson correlation between the skewness of 95\% confidence intervals and the skewness of left-skewed skew-normal true effects. Scenarios with no true heterogeneity are omitted, since then all true effects equal the true mean effect. Error bars represent Monte Carlo standard errors (MCSE).}
\label{fig:ciskessn}
\end{figure}
\end{landscape}

\begin{landscape}
\begin{figure}
\centering
\begin{knitrout}
\definecolor{shadecolor}{rgb}{0.969, 0.969, 0.969}\color{fgcolor}

{\centering \includegraphics[width=1\linewidth]{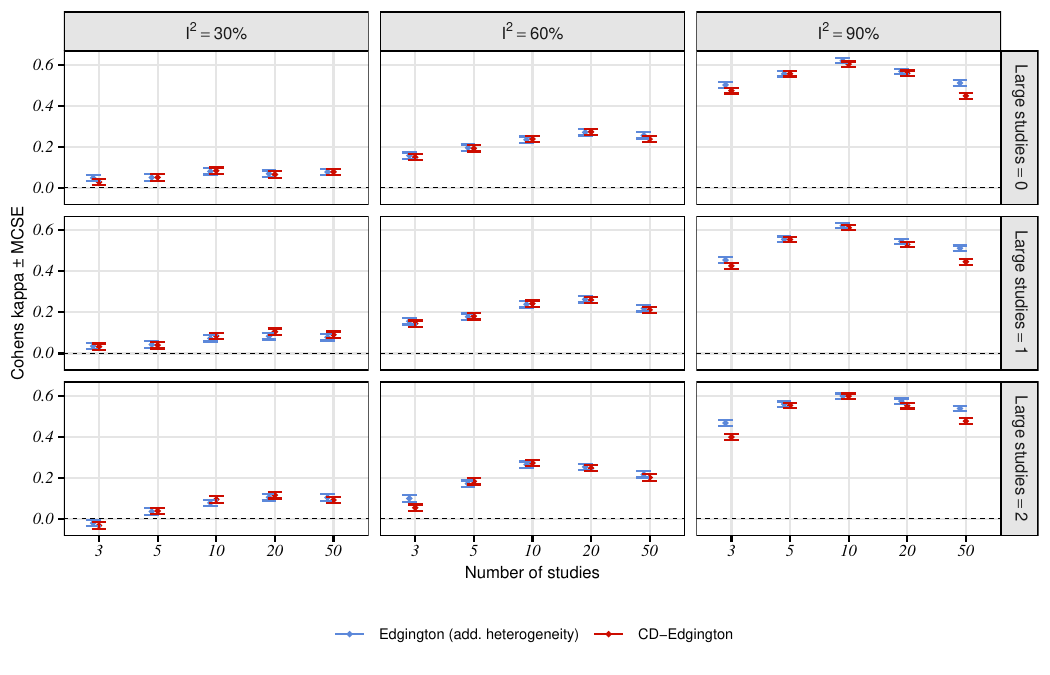} 

}

\end{knitrout}
\caption{Cohens kappa for sign agreement between the skewness of 95\% confidence intervals and the skewness of true effects distributed according to a normal distribution. Error bars represent Monte Carlo standard errors (MCSE).}
\label{fig:cikappanores}
\end{figure}
\end{landscape}

\begin{landscape}
\begin{figure}
\centering
\begin{knitrout}
\definecolor{shadecolor}{rgb}{0.969, 0.969, 0.969}\color{fgcolor}

{\centering \includegraphics[width=1\linewidth]{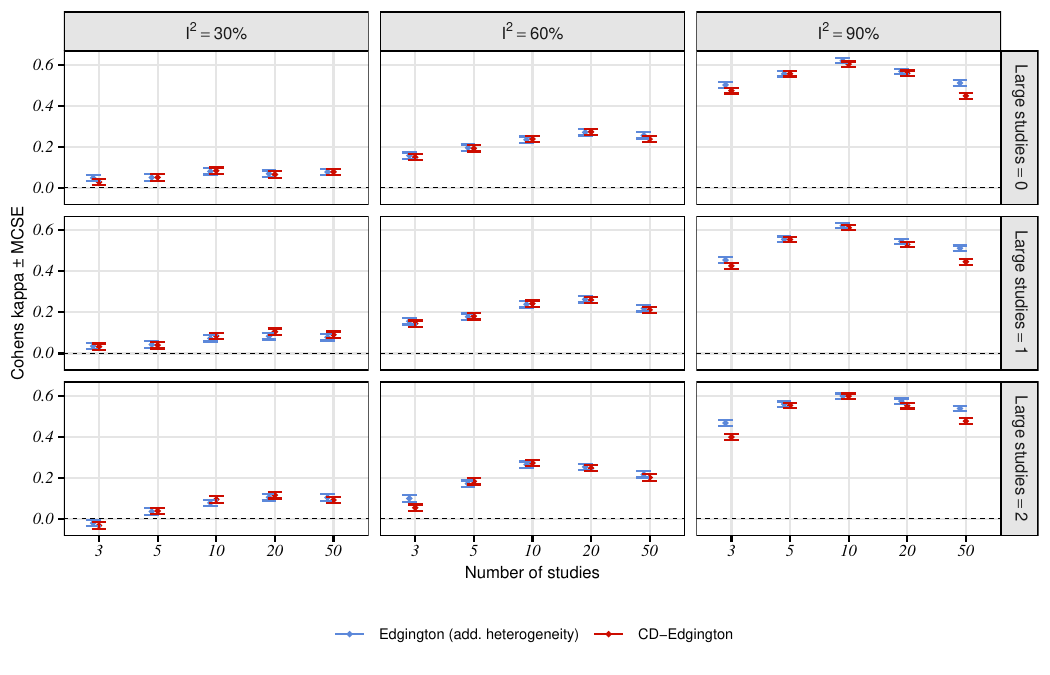} 

}

\end{knitrout}
\caption{Cohens kappa for sign agreement between the skewness of 95\% confidence intervals and the skewness of true effects distributed according to a left-skewed skew-normal distribution. Error bars represent Monte Carlo standard errors (MCSE).}
\label{fig:cikappasnes}
\end{figure}
\end{landscape}

\begin{landscape}
\begin{figure}
\centering
\begin{knitrout}
\definecolor{shadecolor}{rgb}{0.969, 0.969, 0.969}\color{fgcolor}

{\centering \includegraphics[width=1\linewidth]{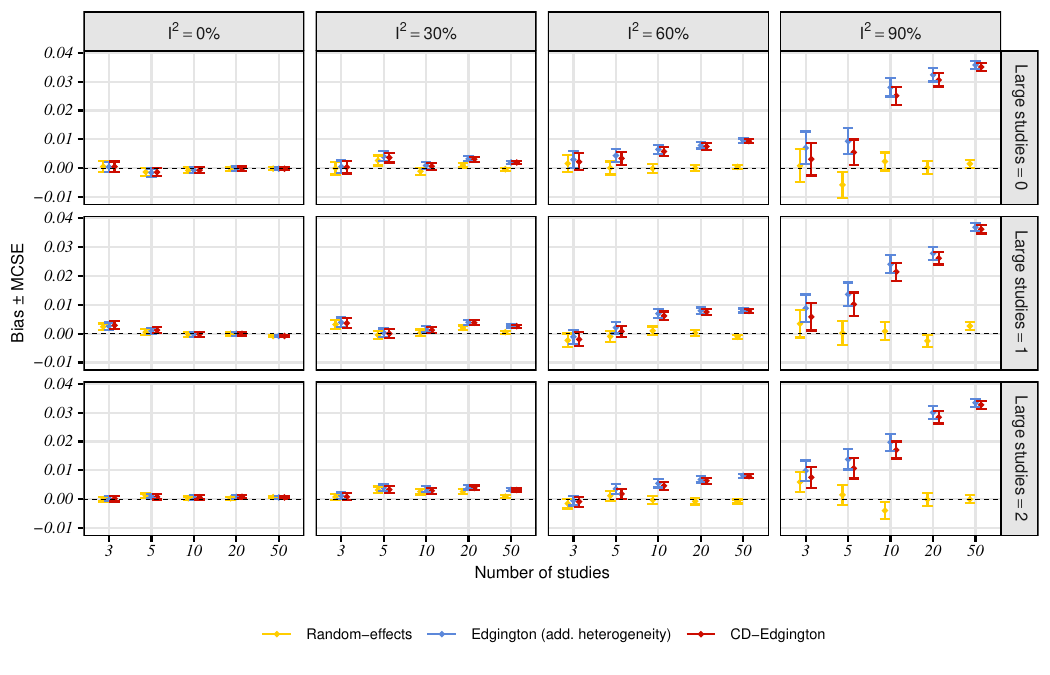} 

}

\end{knitrout}
\caption{Bias for the mean effect, for true effects distributed according to a left-skewed skew-normal distribution. Error bars represent Monte Carlo standard errors (MCSE).}
\label{fig:biassn}
\end{figure}
\end{landscape}

\begin{landscape}
\begin{figure}
\centering
\begin{knitrout}
\definecolor{shadecolor}{rgb}{0.969, 0.969, 0.969}\color{fgcolor}

{\centering \includegraphics[width=1\linewidth]{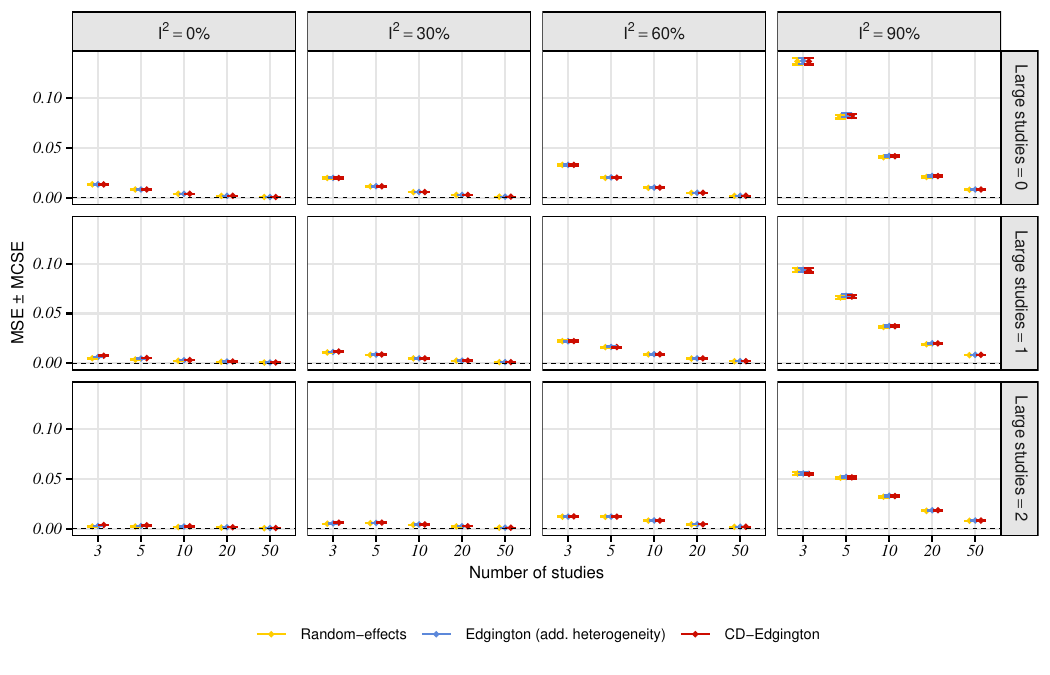} 

}

\end{knitrout}
\caption{Mean squared error (MSE) for the mean effect, for true effects distributed according to a normal distribution. Error bars represent Monte Carlo standard errors (MCSE).}
\label{fig:msenor}
\end{figure}
\end{landscape}

\begin{landscape}
\begin{figure}
\centering
\begin{knitrout}
\definecolor{shadecolor}{rgb}{0.969, 0.969, 0.969}\color{fgcolor}

{\centering \includegraphics[width=1\linewidth]{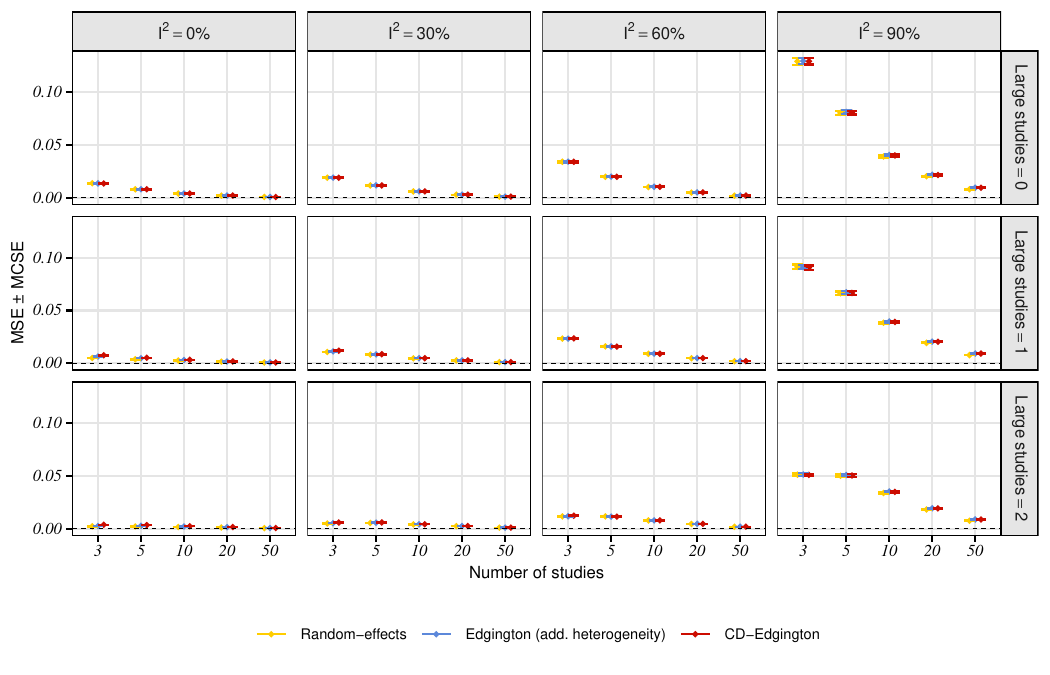} 

}

\end{knitrout}
\caption{Mean squared error (MSE) for the mean effect, for true effects distributed according to a left-skewed skew-normal distribution. Error bars represent Monte Carlo standard errors (MCSE).}
\label{fig:msesn}
\end{figure}
\end{landscape}

\end{document}